\documentclass{article}

\usepackage{graphicx} 
\usepackage{csquotes}
\usepackage{amsmath}
\usepackage{amssymb}
\usepackage{amsfonts}
\usepackage{mathtools}

\usepackage{ifthen}
\newboolean{arxiv}
\setboolean{arxiv}{true}
\ifthenelse{\boolean{arxiv}}{
    \usepackage[style=numeric,maxbibnames=2,maxcitenames=2,backend=biber]{biblatex}
    \addbibresource{bib.bib}
    \AtEveryBibitem{%
    \clearfield{urlyear}%
    }
}{
}

\usepackage[mode=tex]{standalone}
\usepackage{array}
\usepackage{amsfonts}
\usepackage{mathtools}
\usepackage{xintexpr}
\usepackage[usenames,dvipsnames,svgnames]{xcolor}
\usepackage{relsize}
\usepackage{tikz}
\usepackage{tikz-3dplot}
\usepackage{pgfplots}
\usepackage{animate}
\usepackage{graphicx}
\pgfplotsset{compat=1.18} 
\usepgfplotslibrary{external, statistics}
\usepgfplotslibrary{colorbrewer} 
\usepackage{pgfplotstable}
\usepackage{forest}
\usetikzlibrary{positioning, calc, math, matrix}

\usepackage{tikz}

\definecolor{ccecece}{RGB}{206,206,206}
\tikzset{
plane-sphere-o/.pic={
    \path[draw,line width=0.01cm] (0, 0) circle (0.495cm);
  },
plane-sphere-x/.pic={
    \path[draw,line width=0.01cm] (0, 0) circle (0.495cm);
    \path (-0.285, 0.165) -- (0.245, 0.425) -- (0.215, -0.125) -- (-0.245, -0.425) -- cycle;
    \path[draw,fill=ccecece,line cap=round,line join=round,line width=0.01cm]
    (-0.245, 0.425) -- (0.28, 0.16) -- (0.245, -0.425) -- (-0.215, -0.13) -- cycle;
  },
plane-sphere-y/.pic={
    \path[draw,line width=0.01cm] (0, 0) circle (0.495cm);
    \path[draw,fill=ccecece,line cap=round,line join=round,line width=0.01cm]
    (-0.285, 0.165) -- (0.245, 0.425) -- (0.215, -0.125) -- (-0.245, -0.425) -- cycle;
  },
plane-sphere-z/.pic={
    \path[draw,line width=0.01cm] (0, 0) circle (0.495cm);
    \path[draw,fill=ccecece,line cap=round,line join=round,line width=0.01cm]
    (-0.49, 0) -- (0, -0.325) -- (0.485, 0) -- (0, 0.25) -- cycle;
  },
plane-sphere-xy/.pic={
    \path[draw,line width=0.01cm] (0, 0) circle (0.495cm);
    \path[draw,fill=ccecece,line cap=round,line join=round,line width=0.005cm]
    (0, 0.3) -- (0.23, 0.185) -- (0.245, 0.425) -- cycle
    (-0.245, 0.425) -- (-0.235, 0.19) -- (0, 0.3) -- cycle
    (0, 0.3) -- (-0.235, 0.19) -- (-0.285, 0.165) -- (-0.245, -0.425) -- (-0.005, -0.265) -- cycle
    (-0.005, -0.265) -- (0.245, -0.425) -- (0.28, 0.16) -- (0.23, 0.185) -- (0, 0.3) -- cycle;
    \path[draw,line cap=round,line join=round,line width=0.01cm]
    (0.245, 0.425) -- (0, 0.3) -- (0.23, 0.185)
    (-0.235, 0.19) -- (0, 0.3) -- (-0.245, 0.425)
    (0, 0.3) -- (-0.235, 0.19) -- (-0.285, 0.165)
    (-0.245, -0.425) -- (-0.005, -0.265)
    (-0.005, -0.265) -- (0.245, -0.425)
    (0.28, 0.16) -- (0.23, 0.185) -- (0, 0.3);
  },
plane-sphere-xz/.pic={  
    \path[draw,line width=0.01cm] (0, 0) circle (0.495cm);
    \path[draw,fill=ccecece,line cap=round,line join=round,line width=0.005cm]
    (0.275, 0.105) -- (0.26, -0.155) -- (0.485, 0) -- cycle
    (0.26, -0.155) -- (0.045, -0.295) -- (0.245, -0.425) -- cycle
    (-0.23, 0.13) -- (-0.49, 0) -- (0, -0.325) -- (0.045, -0.295) -- (0.26, -0.155) -- cycle
    (0.26, -0.155) -- (0.275, 0.105) -- (0.28, 0.16) -- (-0.245, 0.425) -- (-0.23, 0.13) -- cycle;
    \path[draw,line cap=round,line join=round,line width=0.01cm]
    (0.275, 0.105) -- (0.26, -0.155) -- (0.485, 0)
    (0.245, -0.425) -- (0.26, -0.155) -- (0.045, -0.295)
    (0, -0.325) -- (0.045, -0.295) -- (0.26, -0.155)
    (-0.23, 0.13) -- (-0.49, 0)
    (0.26, -0.155) -- (0.275, 0.105) -- (0.28, 0.16)
    (-0.245, 0.425) -- (-0.23, 0.13);
  },
plane-sphere-yz/.pic={
    \path[draw,line width=0.01cm] (0, 0) circle (0.495cm);
    \path[draw,fill=ccecece,line cap=round,line join=round,line width=0.005cm]
    (-0.265, -0.15) -- (-0.28, 0.105) -- (-0.49, 0) -- cycle
    (-0.05, -0.295) -- (-0.265, -0.15) -- (-0.245, -0.425) -- cycle
    (-0.265, -0.15) -- (-0.05, -0.295) -- (0, -0.325) -- (0.485, 0) -- (0.225, 0.13) -- cycle
    (0.225, 0.13) -- (0.245, 0.425) -- (-0.285, 0.165) -- (-0.28, 0.105) -- (-0.265, -0.15) -- cycle;
    \path[draw,line cap=round,line join=round,line width=0.01cm]
    (-0.49, 0) -- (-0.265, -0.15) -- (-0.28, 0.105)
    (-0.05, -0.295) -- (-0.265, -0.15) -- (-0.245, -0.425)
    (-0.265, -0.15) -- (-0.05, -0.295) -- (0, -0.325)
    (0.485, 0) -- (0.225, 0.13)
    (-0.285, 0.165) -- (-0.28, 0.105) -- (-0.265, -0.15)
    (0.225, 0.13) -- (0.245, 0.425);
  },
plane-sphere-xyz/.pic={
    \path[draw,line width=0.01cm] (0, 0) circle (0.495cm);
    \path[draw,fill=ccecece,line cap=round,line join=round,line width=0.005cm]
    (-0.245, 0.425) -- (-0.235, 0.19) -- (0, 0.3) -- cycle
    (0, 0) -- (0, 0) -- (-0.265, -0.15) -- (-0.05, -0.295) -- (0, -0.325) -- (0.045, -0.295) -- (0.045, -0.295) -- (0.26, -0.155) -- cycle
    (0, 0.3) -- (0, 0) -- (0, 0) -- (0.26, -0.155) -- (0.275, 0.105) -- (0.28, 0.16) -- (0.23, 0.185) -- cycle
    (0, 0.3) -- (-0.235, 0.19) -- (-0.285, 0.165) -- (-0.28, 0.105) -- (-0.265, -0.15) -- (0, 0) -- (0, 0) -- cycle
    (0.045, -0.295) -- (0.045, -0.295) -- (0.245, -0.425) -- (0.26, -0.155) -- cycle
    (0.26, -0.155) -- (0.485, 0) -- (0.275, 0.105) -- cycle
    (-0.05, -0.295) -- (-0.265, -0.15) -- (-0.05, -0.295) -- cycle
    (-0.265, -0.15) -- (-0.28, 0.105) -- (-0.49, 0) -- cycle
    (0.23, 0.185) -- (0.245, 0.425) -- (0, 0.3) -- cycle
    (-0.05, -0.295) -- (-0.265, -0.15) -- (-0.245, -0.425) -- cycle;
    \path[draw,line cap=round,line join=round,line width=0.01cm]
    (-0.235, 0.19) -- (0, 0.3) -- (-0.245, 0.425)
    (-0.265, -0.15) -- (-0.05, -0.295) -- (0, -0.325) -- (0.045, -0.295) -- (0.045, -0.295) -- (0.26, -0.155)
    (0.26, -0.155) -- (0.275, 0.105) -- (0.28, 0.16) -- (0.23, 0.185) -- (0, 0.3)
    (0, 0.3) -- (-0.235, 0.19) -- (-0.285, 0.165) -- (-0.28, 0.105) -- (-0.265, -0.15)
    (0.245, -0.425) -- (0.26, -0.155) -- (0.045, -0.295) -- (0.045, -0.295)
    (0.275, 0.105) -- (0.26, -0.155) -- (0.485, 0)
    (-0.05, -0.295) -- (-0.265, -0.15) -- (-0.05, -0.295) -- cycle
    (-0.49, 0) -- (-0.265, -0.15) -- (-0.28, 0.105)
    (0.245, 0.425) -- (0, 0.3) -- (0.23, 0.185)
    (-0.05, -0.295) -- (-0.265, -0.15) -- (-0.245, -0.425);
  },
}

\usepackage[mode=text,group-four-digits=true,group-separator={,},exponent-product=\cdot]{siunitx}
\usepgfplotslibrary{units} 

\providecommand{\complexity}[1]{\ensuremath{\mathcal{O}(#1)}}
\providecommand{\positived}{\ensuremath{{\mathbb{N}_0}^d}}
\providecommand{\level}{\ensuremath{\vec{\ell}}}
\providecommand{\indexd}{\ensuremath{\vec{i}}}
\providecommand{\variance}{\ensuremath{\mathbb{V}}}
\providecommand{\expectation}{\ensuremath{\mathbb{E}}}
\providecommand{\thingitenk}{\texttt{Thingi10K}}
\providecommand{\stringify}[1]{\ensuremath{\$(#1)}}
\providecommand{\lambdify}[1]{\ensuremath{\hat{\$}(#1)}}
\providecommand{\stringifyInDimension}[2]{\ensuremath{\$^{#1}(#2)}}
\providecommand{\lambdifyInDimension}[2]{\ensuremath{\hat{\$}^{#1}(#2)}}

\DeclarePairedDelimiter{\norm}{\lVert}{\rVert}
\DeclarePairedDelimiter{\abs}{\lvert}{\rvert}

\tikzset{
  octree_t/.style={
    orange,
  },
  omnione_t/.style={
    teal,
  },
}
\pgfplotsset{
  discard if not/.style 2 args={
    filter discard warning=false,
    x filter/.append code={
      \edef\tempa{\thisrow{#1}}%
      \edef\tempb{#2}%
      \ifnum\pdfstrcmp{\tempa}{\tempb}=0
      \else
          \def\pgfmathresult{inf}%
      \fi
    }
  },
  compat=1.18,
  table/col sep=comma,
  table/search path={.,./data/,../data/},
  cycle list name=exotic,
  powtwoxlabels/.style={
    log basis x=2,
		xticklabel style = {align=center, font=\small, rotate=60},
    xticklabel={\xinttheiexpr[0]2^\tick\relax},
  },
  powtenylabels/.style={
    log basis y=10,
		yticklabel style = {align=center, font=\small, rotate=60},
    yticklabel={\xinttheiexpr[0]10^\tick\relax},
  },
  log x ticks with fixed point/.style={
      xticklabel={
        \pgfkeys{/pgf/fpu=true}
        \pgfmathparse{exp(\tick)}%
        \pgfmathprintnumber[fixed relative, precision=3]{\pgfmathresult}
        \pgfkeys{/pgf/fpu=false}
      }
  },
  scalingplot/.style={
    grid=both,
    table/x=input-size,
    legend pos=north west,
    xmin=1,
    xmax=2097152,
    powtwoxlabels,
  },
  errorplot/.style={
		ylabel = {$L_1$ error (Monte Carlo)},
		xlabel = {\shortstack{$N$ \\ Allowed number of cuboids in $\Omega = (0,1)^3$}},
  },
  errorplot4d/.style={
    errorplot,
    xlabel = {\shortstack{$N$ \\ Allowed number of rectangles in $\Omega = (0,1)^4$}},
  },
  linear/.style={
    domain=1:262144,
    samples=2,
    mark=none,
    solid,
    opacity=0.7,
    pink,
  },
  octree/.style={
    octree_t,every mark/.append style={fill=orange!80!black,solid},mark size=1.5pt,mark=square*
  },
  omnione/.style={
    omnione_t,every mark/.append style={fill=teal!80!black,solid},mark=diamond*
  },
  octree filtered/.style={
    discard if not={tree}{octree},
    octree,
  },
  omnione filtered/.style={
    discard if not={tree}{omnitree_1},
    omnione,
  },
  thingitet filtered/.style={
    discard if not={thingi_file_id}{0},
    thingitet,
  },
  thingisphere filtered/.style={
    discard if not={thingi_file_id}{1},
    thingisphere,
  },
  thingirod filtered/.style={
    discard if not={thingi_file_id}{2},
    thingirod,
  },
  thingihilbert filtered/.style={
    discard if not={thingi_file_id}{53750},
    thingihilbert,
  },
  thingikitty filtered/.style={
    discard if not={thingi_file_id}{100349},
    thingikitty,
  },
  thingicube filtered/.style={
    discard if not={thingi_file_id}{187279},
    thingicube,
  },
  thingigear filtered/.style={
    discard if not={thingi_file_id}{99905},
    thingigear,
  },
  minmaxbars/.style={
    error bars/y dir=both,
    error bars/y explicit,
    error bars/error bar style={solid},
    table/y expr=\thisrow{l1error_mean},
    table/y error plus expr=\thisrow{l1error_max}-\thisrow{l1error_mean},
    table/y error minus expr=\thisrow{l1error_mean}-\thisrow{l1error_min},
  },
}
\tikzset{
  thingitet/.style={
    solid,
    color=Dark2-A,
  },
  thingisphere/.style={
    densely dashdotted,
    color=Dark2-B,
  },
  thingirod/.style={
    densely dashdotdotted,
    color=Dark2-C,
  },
  thingihilbert/.style={
    dashed,
    color=Dark2-D,
  },
  thingikitty/.style={
    densely dashed,
    color=Dark2-E,
  },
  thingicube/.style={
    dotted,
    color=Dark2-H,
  },
  thingigear/.style={
    color=Dark2-F,
  },
  thingiplane/.style={
    thick,
    color=Dark2-F,
  },
  thingilabel/.style={
    font=\small,
    anchor=west,
    xshift=0.5ex,
    inner sep=0.7ex,
  },
  thingilabel_up/.style={
    thingilabel,
    anchor=south,
    xshift=-0.5ex,
  },
  thingilabel_left/.style={
    thingilabel,
    anchor=east,
    xshift=-0.5ex,
  },
}

\usepackage[mode=text,group-four-digits=true,group-separator={,},exponent-product=\cdot]{siunitx}
\usepackage{subcaption}
\usepackage{algorithm}
\usepackage{algpseudocode}

\pgfplotsset{compat=1.18}
\graphicspath{{./gfx/}}

\usepackage{hyperref}
\usepackage[capitalize,nameinlink]{cleveref}
\usepackage{amsthm}
\newtheorem{definition}{Definition}[section]

\ifthenelse{\boolean{arxiv}}{
\title{The Beauty of Anisotropic Mesh Refinement: \\ Omnitrees for Efficient Dyadic Discretizations}
\author{Theresa Pollinger \and Masado Ishii \and Jens Domke}
\date{RIKEN Center for Computational Science}
}{
\journal{Mathematics and Computers in Simulation}
}

\begin{document}
\ifthenelse{\boolean{arxiv}}{
\maketitle
}{
\begin{frontmatter}
\title{The Beauty of Anisotropic Mesh Refinement: Omnitrees for Efficient Dyadic Discretizations}
\author{Theresa Pollinger\corref{cor1}}
\author{Masado Ishii}
\author{Jens Domke}
\affiliation{organization={RIKEN Center for Computational Science},
            addressline={7-1-26 Minatojima-minami-machi}, 
            city={Kobe},
            postcode={650-0047}, 
            state={Hyogo},
            country={Japan}}
\cortext[cor1]{Corresponding author}
}
\begin{abstract}
    Structured adaptive mesh refinement (AMR), commonly implemented via quadtrees and octrees, underpins a wide range of applications including databases, computer graphics, physics simulations, and machine learning.
    However, octrees enforce isotropic refinement in regions of interest, which can be especially inefficient for problems that are intrinsically anisotropic---much resolution is spent where little information is gained.
    This paper presents omnitrees as an anisotropic generalization of octrees and related data structures.
    Omnitrees allow to refine only the locally most important dimensions, providing tree structures that are less deep than bintrees and less wide than octrees.
    As a result, the convergence of the AMR schemes can be increased by up to a factor of the dimensionality $d$ for very anisotropic problems, quickly offsetting their modest increase in storage overhead.
    We validate this finding on the problem of binary shape representation across \num{4166} three-dimensional objects:
    Omnitrees increase the mean convergence rate by \num{1.5}${\times}$, require less storage to achieve equivalent error bounds, and maximize the information density of the stored function faster than octrees.{}
    These advantages are projected to be even stronger for higher-dimensional problems.
    We provide a first validation by introducing a time-dependent rotation to create four-dimensional representations, and discuss the properties of their $4$-d octree and omnitree approximations.
    Overall, omnitree discretizations can make existing AMR approaches more efficient, and open up new possibilities for high-dimensional applications.
\end{abstract}

\ifthenelse{\boolean{arxiv}}{}{
\begin{keyword}
adaptive mesh refinement \sep octrees \sep anisotropy \sep sensitivity analysis \sep information theory

\end{keyword}
\end{frontmatter}
}

\section{Introduction}\label{sec:intro} 

Adaptively resolved data structures have served as a useful abstraction in diverse areas of computing.
As a main example, octrees~\cite{meagherGeometricModelingUsing1982,sametQuadtreeRelatedHierarchical1984} (and quadtrees as their two-dimensional siblings), were formalized in the 1980s and have found applications not only in computer graphics and spatial modeling~\cite{meagherGeometricModelingUsing1982,whangOctreeRAdaptiveOctree1995}, but also in particle methods~\cite{greengardFastAlgorithmParticle1987,ambrosianoFastMultipoleMethod1988}, simulation codes for fluid dynamics~\cite{losassoSpatiallyAdaptiveTechniques2006,mirzadehParallelLevelsetMethods2016,judeOctreebasedCartesianNavier2022}, astrophysics~\cite{nakasatoAstrophysicalParticleSimulations2012,daissPizDaintStars2019a}, and many other scientific fields, as well as spatial databases and nearest-neighbor search~\cite{tuEtreeDatabaseorientedMethod2004,hornungOctoMapEfficientProbabilistic2013,schonOctreebasedIndexing3D2013,chenParallelNNParallelOctreebased2023}, and more recently, transformer-based machine-learning architectures~\cite{dengEfficientAutoregressiveShape2025a,zhangAdaptivePatchingHighresolution2024a}.
Octrees can significantly decrease the computational and storage costs in all these applications when compared to storing all data at uniform resolution.
Combined with space-filling curves~\cite{baderSpaceFillingCurves2013}, they furthermore favor cache efficiency in applications where data is exchanged locally, and allow for natural parallelization~\cite{minyardOctreePartitioningHybrid1998,bursteddeP4estScalableAlgorithms2011,holkeT8codeModularAdaptive2025}.{}
As a result, the term \enquote{octree} is nearly synonymous with structured Adaptive Mesh Refinement (AMR) in the scientific computing literature.

\begin{figure}
    \centering
    \begin{subfigure}[t]{0.25\textwidth}
        \centering
        \hspace*{-3em}\includestandalone[width=\textwidth]{gfx/cuboids_solid}
        \caption{$3$-d cube omnitree discretization}
        \label{fig:cuboid:cube}
    \end{subfigure}\hfill
    \begin{subfigure}[t]{0.45\textwidth}
        \centering
        \hspace*{-5em}\includestandalone[mode=tex,width=1.2\textwidth]{gfx/cuboids_tree_animated}
        \caption{\textellipsis and corresponding tree
        ("$<$" and "$>$" toggle glyphs with \href{https://tex.stackexchange.com/questions/235139/using-the-animate-package-without-adobe}{various pdf readers})
        }
        \label{fig:cuboid:tree}
    \end{subfigure}\hfill
    \begin{subfigure}[t]{0.25\textwidth}
        \centering
        \hspace*{-3em}\includestandalone[width=\textwidth]{gfx/cuboids_octree}
        \caption{Matching octree discretization}
        \label{fig:cuboid:oct}
    \end{subfigure} 
    \caption{
       The left image shows the unit cube discretized with omnitree and Z order ($d=3$).
       The cube is resolved with different anisotropic resolutions in different parts of the cube.
       This is also reflected in the tree representation, given in the middle.
       To the right, the cube is discretized with an octree; 
       Instead of 14 cuboids, the octree requires 57 cubes to represent the same scales as the omnitree.
       (The resulting tree becomes too wide to be printed on this page.)
       The colors denote the location in storage, similar colors will be stored closer to each other. One can observe that the Z order results in a certain degree of data locality.
    }
    \label{fig:cuboid}
\end{figure}

Despite their widespread utility, there are areas of computational science and engineering where the use of octrees becomes less practical.
For instance, high-fidelity plasma micro-turbulence simulations exhibit both high dimensionalities (five to six dimensions in phase space, plus time~\cite{brizardFoundationsNonlinearGyrokinetic2007a}) and fine-grained anisotropic structures (\enquote{filamentation}~\cite{ghizzoLowHighfrequencyNature2020}) carrying significant energy, and thus they should be resolved for realistic results.
Octrees, which refine isotropically, exhibit the \enquote{curse of dimensionality}: 
In a $6$-d discretization, the number of refined cells explodes by a factor of $64=2^6$ per refinement, requiring significant memory and compute resources.
As a result, AMR has not found widespread use in this field.

This paper presents \emph{omnitrees} as a generalization of octrees with superior adaptivity.
An omnitree can refine a region in several dimensions, or in just one, as necessary.
This ability does not merely improve efficiency for many $3$-d problems---it cuts through the curse of dimensionality, making structured AMR tractable for anisotropy in higher dimensions.

We introduce a formalism for omnitrees (\cref{sec:overview}) and compare it to existing approaches (\cref{sec:related}), most of which can be realized as special cases of omnitrees in our formulation.{}
\Cref{sec:refinement} further analyzes the tree refinement algorithm, and shows that omnitrees increase the order of convergence by up to a factor of $d$ compared to octrees, depending on the anisotropy of the approximated problem (but independent of the method used).
In \cref{sec:eval}, we describe our evaluation method, the binary representation of three- and four-dimensional shapes, the results of which we evaluate in \cref{sec:results}.
\Cref{sec:discussion} discusses how these findings translate to more practical science applications, and what problems will need to be addressed to achieve this.

\section{Omnitrees for Adaptive Dyadic Refinement}\label{sec:overview}

This section briefly introduces the concept of omnitrees, to allow for a comparison with related data structures.
Algorithmic details will be further discussed in \cref{sec:refinement}.

Omnitrees, like octrees, are designed to partition a $d$-dimensional Cartesian domain $\Omega \subset \mathbb{R}^d$ into $N$ non-overlapping subdomains,
\begin{equation}
\overline \Omega = \bigcup_{1 \leq i \leq N} \overline{Q_i}, \quad \text{and} \quad Q_i \cap Q_j = \emptyset \quad \text{for } i \ne j,
\end{equation}
where the overline denotes the closure of $d$-dimensional hyper-rectangles $Q = (x_1^-,x_1^+) \times \dots \times (x_d^-,x_d^+)$.
(For brevity, we will write \enquote{rectangles}, or, when emphasizing $3$-d, \enquote{cuboids}.)
We suppose the dimensions are indexed by the set $\mathcal{D}$, with $\lvert\mathcal{D}\rvert=d$.
From here, we use the unit hypercube as the domain, $\Omega = (0,1)^\mathcal{D} $, without loss of generality.
An omnitree describes a hierarchical \textit{dyadic} partition of the unit hypercube, meaning that partition boundaries always occur at multiples of appropriate powers of $\frac{1}{2}$.


\begin{definition}[Omnitree]\label{def:omnitree}
  A $d$-dimensional omnitree is a rooted tree with the following additional structure:
    \begin{enumerate}
        \item A function $\sigma$ that picks a subset of dimensions to be split at any node, $v \mapsto \sigma(v) \subset \mathcal{D}$.
        Equivalently, $\sigma$ is represented by its indicator map, which is a $d$-digit binary label $\vec{b}(v)\colon\: b_j=1 \leftrightarrow j \in \sigma(v)$ (see \cref{fig:splits}).
        \label{item:label}
        \item Every nonleaf node $v$ has $2^{d'}$ children, where $d' = \#\sigma(v) \leq d$ is the number of dimensions bisected at $v$ and thus the number of $1$s in its label $\vec{b}(v)$.
        The children are enumerated by an index set $\kappa(v) = \{0,1\}^{\sigma(v)}$, where an index $e \in \kappa(v)$ is a named binary tuple, $(e_j \in \{0,1\})_{j \in \sigma(v)}$, that refers to the intersection of half spaces in the split dimensions.
        ($e_j$ is undefined if $j \not \in \sigma(v)$.)
        \label{item:numchildren}
    \end{enumerate}
\end{definition}

For uniqueness, we stipulate a third condition:
\begin{definition}[Normalized Omnitree]\label{def:normalized_omnitree}
    \begin{enumerate}
        \addtocounter{enumi}{2}
        \item In any label position $j$, a node must have $b_j = 1$ if all its children have $b_j=1$.
        \label{item:unique}
    \end{enumerate}
\end{definition}

\begin{figure*}
    \centering
    \begin{subfigure}{0.17\textwidth}
        \centering
        \begin{tikzpicture}[scale=1.0]
    \node[] at (-0.8,0) {000};
    \pic [scale=0.8] at (0,0) {plane-sphere-o};
\end{tikzpicture}
        \caption{}
        \label{fig:splits:full}
    \end{subfigure}\hspace*{1em}
    \begin{subfigure}{0.17\textwidth}
        \centering
        \begin{tikzpicture}[scale=1.0]
    \node[] at (-0.8,0) {111};
    \pic [scale=0.8] at (0,0) {plane-sphere-xyz};
\end{tikzpicture}
        \caption{}
        \label{fig:splits:xyz}
    \end{subfigure}\\
    \begin{subfigure}{0.15\textwidth}
        \centering
        \begin{tikzpicture}[scale=1.0]
    \node[] at (-0.8,0) {100};
    \pic [scale=0.8] at (0,0) {plane-sphere-x};
\end{tikzpicture}
        \caption{}
        \label{fig:splits:x}
    \end{subfigure}\hfill
    \begin{subfigure}{0.15\textwidth}
        \centering
        \begin{tikzpicture}[scale=1.0]
    \node[] at (-0.8,0) {010};
    \pic [scale=0.8] at (0,0) {plane-sphere-y};
\end{tikzpicture}
        \caption{}
        \label{fig:splits:y}
    \end{subfigure}\hfill
    \begin{subfigure}{0.15\textwidth}
        \centering
        \begin{tikzpicture}[scale=1.0]
    \node[] at (-0.8,0) {001};
    \pic [scale=0.8] at (0,0) {plane-sphere-z};
\end{tikzpicture}
        \caption{}
        \label{fig:splits:z}
    \end{subfigure}\hfill
    \begin{subfigure}{0.15\textwidth}
        \centering
        \begin{tikzpicture}[scale=1.0]
    \node[] at (-0.8,0) {110};
    \pic [scale=0.8] at (0,0) {plane-sphere-xy};
\end{tikzpicture}
        \caption{}
        \label{fig:splits:xy}
    \end{subfigure}\hfill
    \begin{subfigure}{0.15\textwidth}
        \centering
        \begin{tikzpicture}[scale=1.0]
    \node[] at (-0.8,0) {011};
    \pic [scale=0.8] at (0,0) {plane-sphere-yz};
\end{tikzpicture}
        \caption{}
        \label{fig:splits:yz}
    \end{subfigure}\hfill
    \begin{subfigure}{0.15\textwidth}
        \centering
        \begin{tikzpicture}[scale=1.0]
    \node[] at (-0.8,0) {101};
    \pic [scale=0.8] at (0,0) {plane-sphere-xz};
\end{tikzpicture}
        \caption{}
        \label{fig:splits:xz}
    \end{subfigure}
    \caption{
       Correspondence between tree labels and spatial refinement:
        The labels indicate in which direction(s) a node is refined, \enquote{1} means refinement in the respective dimension and \enquote{0} means no refinement.
        For example, \enquote{100} in (c) denotes refinement in the first dimension, \enquote{011} in (g) gives the refinements perpendicular to the first dimension.
        In the case of no refinement (\enquote{000}, \cref{fig:splits:full}), the node is a leaf in the tree, indicated by the circle.
        The isotropic refinements (a)-(b) (\enquote{000} and \enquote{111}) are the ones that can also be represented in an octree; the anisotropic refinements (c)-(h) are exclusive to omnitrees.
    }
    \label{fig:splits}
\end{figure*}

We designate a unique omnitree structure via the normalization property, which is preserved by our refinement algorithm.
Unnormalized omnitrees can partition a given set of non-overlapping rectangles in various ways at different levels of the tree.
Normalization greedily achieves a shallowest tree by gathering splits $(b_j = 1)$ as close to the root node as possible, thus serving as a type of unique minimum over general omnitrees.

The association of each node $v$ to a rectangle $Q$ is denoted $\textrm{rect}(v) = Q$.
The largest rectangle, $\Omega$, is assigned to the root node.
The rectangle of a node $v$ is bisected in dimensions $\sigma(v)$ and the resulting sub-rectangles are assigned to its $2^{d'(v)}$ children.
Thus $\textrm{rect}(v)$ depends only on $\Omega$ and the binary labels between $v$ and the root.

In practice, we encode the whole tree structure and spatial partition as a flat sequence of the binary labels, $\vec{b}(v)$.
The functions $\sigma(v)$ and $\kappa(v)$ are defined only for later notational convenience and need not be computed or stored in memory.
Condition~\ref{item:numchildren} implies that a node with an all-zero label ($\vec{b}=\vec{0}$) would be redundantly assigned the same rectangle as its only child.
Disallowing this case, we make a special convention to interpret $\vec{0}$ labels as leaf nodes.
\Cref{fig:cuboid:cube,fig:cuboid:tree} show a tree of binary labels representing a possible dyadic decomposition of the unit hypercube.

To linearize the tree, we require a sequencing of each set of child indices $\kappa(v).$
For the purposes of this paper, we assume the standard Z-curve order~\cite{baderSpaceFillingCurves2013,haverkortSixteenSpacefillingCurves2018a}.
Due to the dimension-wise separability of the Z-curve, it is readily applied to omnitrees, resulting in a sequence of leaf nodes and, ultimately, of their $N$ associated rectangles:
\begin{equation}
  t: i \in \{1,\dots,N\} \mapsto Q_i \subset \Omega.
\end{equation}

Its inverse is an index map on $d$-dimensional coordinates:
\begin{equation}
  p: x \in \Omega \mapsto i : x \in t(i).
\end{equation}

This allows us to store any omnitree-discretized function $g':\Omega \to \mathbb{R}$ in a flat vector data structure $\hat{g}$,
\begin{equation}
  \label{eq:discrete-function}
    g'(x) = \hat{g}[p(x)]
\end{equation}
where $\mathrm{length}(\hat{g})=N$ and the square brackets denote a data structure lookup operation.

Now that we have defined omnitree discretizations, in the next section we compare them with other hierarchical discretizations.
The procedure to grow an omnitree by successive refinements is covered in \cref{sec:refinement}.

\section{Related Work}\label{sec:related}

In the 2000s and 2010s, there were several lines of work developing and using omnitrees for aerospace simulation:
First, Domel et al.~\cite{domelSplitflowProgress3D2000,baruzziniFundamentalChallengesMicroVanes2009,domelTimeTensorRapidConvergence2010,domelPropulsionAerodynamicsWorkshop2019} introduced the concept of anisotropic Cartesian discretizations and applied it to thermal-fluid simulations. 
They also first used the term \enquote{Omni-tree} for this type of data structure.
Second, Ogawa~\cite{ogawaAdaptiveCartesianMesh2003,ogawaParallelizationAdaptiveCartesian2003,ogawaNumericalAlgorithmBased2009} introduced a flow solver on omnitrees and illustrated its straightforward parallelization to multiple cores.
Third, Sang et al.~\cite{sangOmnitreeAdaptiveCartesian2003,sangApplicationOmnitreeAdaptive2003,sangNumericallyAnalyzingMore2011,sangComparisonOctreeOmniTree2013} also developed and analyzed flow solvers operating on omnitree data.
In addition, the commercial software CFD-VisCART (distributed by ESI Group) implements omnitree discretizations for fluid simulations.
These have been used successfully to investigate a range of cardiovascular engineering applications~\cite{vardakisUsingMulticompartmentalPoroelasticity2021,lyuNovelComputationalPreProcedural2024,peachVirtualComparisonECLIPs2019}.

These existing works applied omnitree concepts directly to the domain of CFD simulations but were restricted to two and three spatial dimensions.
Furthermore, as the emphasis was on applications, the data structures were not described in detail with respect to memory management, which is nevertheless essential to predict performance on contemporary high-performance computer architectures.
The current work contributes an information-theoretical analysis of omnitrees, the extension to arbitrary spatial dimensions, and an explicit description of our proposed memory layout of linearized omnitree binary data.

Apart from the octree~\cite{sametQuadtreeRelatedHierarchical1984}, which will be the baseline for comparison in the remainder of the paper, other prominent dyadic discretizations are roughly categorized as either $k$-d trees or Adaptive Multilinear Meshes, described next.

Classically, $k$-d trees~\cite{bentleyMultidimensionalBinarySearch1975} split their domain into two subdomains at every parent tree node.
For multiple dimensions, they typically iterate all dimensions in a round-robin fashion for successive splittings.
Bintrees~\cite{knowltonProgressiveTransmissionGreyscale1980,sametQuadtreeRelatedHierarchical1984}, a special case of $k$-d trees, always bisect in the geometric middle, rather than at data-dependent points.
An octree can always be embedded in a bintree, which will be $d$ times as deep for representing the same data.
While anisotropy can be represented in a bintree, this is only to a very limited extent: The achievable aspect ratio of every rectangle can be at most $2:1$.
All bintree discretizations can be represented as omnitrees, and as omnitrees will usually have much lower depth.
This is especially true if condition~3 in \cref{def:normalized_omnitree} is fulfilled.
Compared to $k$-d trees (narrow but deep) and octrees (shallow but wide), omnitrees are more flexible, as they are able to minimize both tree depth and width simultaneously.{}



A recent approach, Adaptive Multilinear Meshes (AMM)~\cite{bhatiaAMMAdaptiveMultilinear2022}, introduces a two- to three-dimensional solution to anisotropic dyadic adaptivity for data streaming and compression for piecewise multilinear functions.
In addition to providing the different anisotropic refinements as illustrated in \cref{fig:splits}, they implement a framework for incrementally updating the data, even if it is only partly known.
Like omnitrees in this work, AMM uses a pointerless representation of the tree data structure, and shares the concept of an integer location encoding of nodes.
However, whereas we use a linearized bit stream representation of the tree data structure, AMM employs a hash map using the individual location codes as hashes, which results in a much larger memory footprint for indexing that would be unsuitable to higher dimensions.
At the same time, AMM can omit any explicit storage of leaf node data through the use of wavelet functions on all parent nodes, which in turn can be an efficient way of saving memory.
The wavelet methods described in the AMM paper should be applicable to omnitrees as well.
In this case, the same functions are representable on both AMM and omnitrees.

\section{Refining an Omnitree}\label{sec:refinement}

\begin{figure*}
    \begin{subfigure}[b]{0.3\textwidth}
        \centering
        \begin{tikzpicture}
\draw[very thin, white] (0,0) rectangle (1,1);
\definecolor{color_0}{RGB}{8,92,248}
\definecolor{color_1}{RGB}{103,175,29}
\definecolor{color_2}{RGB}{244,204,31}
\definecolor{color_3}{RGB}{252,105,156}
\draw[very thin, gray, fill=color_0,fill opacity=0.3] (0.000000,0.000000) rectangle (0.500000,0.500000);
\node[inner sep=0] (A) at (0.25,0.25) {};
\draw[very thin, gray, fill=color_1,fill opacity=0.3] (0.000000,0.500000) rectangle (0.500000,1.000000);
\node[inner sep=0] (B) at (0.25,0.75) {};
\draw[very thin, gray, fill=color_2,fill opacity=0.3] (0.500000,0.000000) rectangle (0.750000,1.000000);
\node[inner sep=0] (C) at (0.625,0.5) {};
\draw[very thin, gray, fill=color_3,fill opacity=0.3] (0.750000,0.000000) rectangle (1.000000,1.000000);
\node[inner sep=0] (D) at (0.875,0.5) {};
\draw[thick, densely dotted] (A) -- (B) -- (C) -- (D);
\fill (A) circle (1pt);
\fill (B) circle (1pt);
\fill (C) circle (1pt);
\fill (D) circle (1pt);
\end{tikzpicture}
        \caption{}
        \label{fig:before:square}
    \end{subfigure} 
    \begin{subfigure}[b]{0.37\textwidth}
        \centering
        \hspace*{-2em}\includestandalone[width=\textwidth]{gfx/tree_grandchild_before_split}
        \caption{}
        \label{fig:before:tree}
    \end{subfigure}
    \begin{subfigure}[b]{0.3\textwidth}
        \centering
        \definecolor{color_0}{RGB}{8,92,248}
\definecolor{color_1}{RGB}{103,175,29}
\definecolor{color_2}{RGB}{244,204,31}
\definecolor{color_3}{RGB}{252,105,156}
\begin{tikzpicture}[every node/.style={draw,align=center,text height=2ex,minimum width=2ex, inner sep=0.2ex, fill opacity=0.4, text opacity=1}]
    \matrix [draw=none, matrix of nodes,nodes in empty cells]
    {    10\vphantom{Ag}&
    01\vphantom{Ag}&
    |[fill=color_0]| 00\vphantom{Ag}&
    |[fill=color_1]| 00\vphantom{Ag}&
    10\vphantom{Ag}&
    |[fill=color_2]| 00\vphantom{Ag}&
    |[fill=color_3]| 00\vphantom{Ag}\\
    };
\end{tikzpicture}
        \par\bigskip\bigskip
        \caption{}
        \label{fig:before:desc}
    \end{subfigure}\par\bigskip
    \begin{subfigure}[b]{0.3\textwidth}
        \centering
        \begin{tikzpicture}
\draw[very thin, white] (0,0) rectangle (1,1);
\definecolor{color_0}{RGB}{8,92,248}
\definecolor{color_1}{RGB}{59,158,74}
\definecolor{color_2}{RGB}{153,187,24}
\definecolor{color_3}{RGB}{244,204,31}
\definecolor{color_4}{RGB}{254,144,123}
\definecolor{color_5}{RGB}{246,66,152}
\draw[very thin, gray, fill=color_0,fill opacity=0.3] (0.000000,0.000000) rectangle (0.500000,0.500000);
\node[inner sep=0] (A) at (0.25,0.25) {};
\draw[very thin, gray, fill=color_1,fill opacity=0.3] (0.500000,0.000000) rectangle (0.750000,0.500000);
\node[inner sep=0] (B) at (0.625,0.25) {};
\draw[very thin, gray, fill=color_2,fill opacity=0.3] (0.750000,0.000000) rectangle (1.000000,0.500000);
\node[inner sep=0] (C) at (0.875,0.25) {};
\draw[very thin, gray, fill=color_3,fill opacity=0.3] (0.000000,0.500000) rectangle (0.500000,1.000000);
\node[inner sep=0] (D) at (0.25,0.75) {};
\draw[very thin, gray, fill=color_4,fill opacity=0.3] (0.500000,0.500000) rectangle (0.750000,1.000000);
\node[inner sep=0] (E) at (0.625,0.75) {};
\draw[very thin, gray, fill=color_5,fill opacity=0.3] (0.750000,0.500000) rectangle (1.000000,1.000000);
\node[inner sep=0] (F) at (0.875,0.75) {};
\draw[thick, densely dotted] (A) -- (B) -- (C) -- (D) -- (E) -- (F);
\fill (A) circle (1pt);
\fill (B) circle (1pt);
\fill (C) circle (1pt);
\fill (D) circle (1pt);
\fill (E) circle (1pt);
\fill (F) circle (1pt);
\end{tikzpicture}
        \caption{}
        \label{fig:after:square}
    \end{subfigure}
    \begin{subfigure}[b]{0.37\textwidth}
        \centering
        \hspace*{-2em}\includestandalone[width=\textwidth]{gfx/tree_grandchild_after_split}
        \caption{}
        \label{fig:after:tree}
    \end{subfigure}
    \begin{subfigure}[b]{0.3\textwidth}
        \centering
        \definecolor{color_0}{RGB}{8,92,248}
\definecolor{color_1}{RGB}{59,158,74}
\definecolor{color_2}{RGB}{153,187,24}
\definecolor{color_3}{RGB}{244,204,31}
\definecolor{color_4}{RGB}{254,144,123}
\definecolor{color_5}{RGB}{246,66,152}
\begin{tikzpicture}[every node/.style={draw,align=center,text height=2ex,minimum width=2ex, inner sep=0.2ex, fill opacity=0.4, text opacity=1}]
    \matrix [draw=none, matrix of nodes,nodes in empty cells]
    {    11\vphantom{Ag}&
    |[fill=color_0]| 00\vphantom{Ag}&
    10\vphantom{Ag}&
    |[fill=color_1]| 00\vphantom{Ag}&
    |[fill=color_2]| 00\vphantom{Ag}&
    |[fill=color_3]| 00\vphantom{Ag}&
    10\vphantom{Ag}&
    |[fill=color_4]| 00\vphantom{Ag}&
    |[fill=color_5]| 00\vphantom{Ag}\\
    };
\end{tikzpicture}
        \par\bigskip\bigskip
        \caption{}
        \label{fig:after:desc}
    \end{subfigure}
    \caption{
        Omnitree representation before and after refinement (upper and lower row).
        The two rightmost (orange and pink) rectangles are refined in the vertical direction, which leads to a reordering of nodes in the tree.
        \Cref{fig:before:square,fig:after:square}:
            Block decomposition of the unit square, the Z order is denoted by the dashed line.
        \Cref{fig:before:tree,fig:after:tree}:
            Tree representations, where the black numbers denote node labels, see also \cref{fig:splits}.
            Leaf nodes are circled and colored according to their corresponding rectangle in the square.
        \Cref{fig:before:desc,fig:after:desc}: 
            Pointerless linearized binary descriptor of the omnitree. 
            The dashed line in the tree shows the depth-first traversal used to create the binary descriptor; it matches the Z order in the discretization.
    }
    \label{fig:2d_gradchild_split}
\end{figure*}

While the general idea of omnitrees was introduced in \cref{sec:overview}, the construction and refinement to create valid omnitrees warrants more discussion, which is addressed in this section.

First, we introduce the level-index notation $Q_{\indexd{},\level{}}$ characterizing those rectangles $\subset \Omega = (0,1)^d$ that are compatible with omnitree refinement. In a single dimension, the grid spacing at refinement level $\ell$ is $2^{-\ell}$, yielding intervals indexed by $i$, where $0 \leq i < 2^\ell$. Omnitrees accommodate anisotropic refinement. Thus, in general, the rectangle $Q_{\indexd{},\level{}}$ has per-dimension refinement levels $\level{} \in \positived$, extents given by the tuple $2^{-\level{}}$, and multi-dimensional index
$\indexd{}$ with $i_j \in 0, \ldots, 2^{\ell_j} - 1$ for $j \in 1,\ldots,d $.
The rectangle $Q_{\indexd{},\level}$ occupies the $d$-dimensional space bounded below by $\indexd{}\cdot2^{-\level{}}$ and above by $(\indexd{} + \vec{1})\cdot2^{-\level}$.

\begin{figure*}
    \centering
    \begin{tikzpicture}
    \tikzset{
        horiz/.style={
            inner sep=0,
            scale=0.3,
            text=red,
        },
        vert/.style={
            horiz,
            text=blue,
        },
        interleaved/.style={
            horiz,
            scale=0.8,
            opacity=0.1,
            text=black,
        },
        li/.style={
            horiz,
            scale=0.7,
            text=black,
            align=center,
        },
    }
    
\newcommand{\fancylabel}[4]{%
    {\textcolor{blue}{\ensuremath{\overline{#1}}}%
    \textcolor{red}{\ensuremath{\underline{#2}}}%
    \textcolor{blue}{\ensuremath{\overline{#3}}}%
    \textcolor{red}{\ensuremath{\underline{#4}}}%
    }%
}
\newcommand{\ammcode}[2]{\ensuremath{\left(\begin{smallmatrix}\textcolor{red}{#1}\\\textcolor{blue}{#2}\end{smallmatrix}\right)}}
\newcommand{\smallvector}[2]{\ensuremath{\scriptsize\left[\!\!\begin{smallmatrix}\textcolor{red}{#1}\\\textcolor{blue}{#2}\end{smallmatrix}\!\!\right]}} 
\newcommand{\plh}[0]{\vphantom{1}{\ensuremath{\diamond}}}
\draw[very thin, white] (0,0) rectangle (1,1);
\definecolor{color_0}{RGB}{8,92,248}
\definecolor{color_1}{RGB}{103,175,29}
\definecolor{color_2}{RGB}{244,204,31}
\definecolor{color_3}{RGB}{252,105,156}
\draw[very thin, gray, fill=color_0,fill opacity=0.3] (0.000000,0.000000) rectangle (0.500000,0.500000);
\draw[very thin, gray, fill=color_1,fill opacity=0.3] (0.000000,0.500000) rectangle (0.500000,1.000000);
\draw[very thin, gray, fill=color_2,fill opacity=0.3] (0.500000,0.000000) rectangle (0.750000,1.000000);
\draw[very thin, gray, fill=color_3,fill opacity=0.3] (0.750000,0.000000) rectangle (1.000000,1.000000);
\node[horiz] (h00) at (0.125,-0.125) {$\underline{00}$};
\node[horiz] (h01) at (0.375,-0.125) {$\underline{01}$};
\node[horiz] (h10) at (0.625,-0.125) {$\underline{10}$};
\node[horiz] (h11) at (0.875,-0.125) {$\underline{11}$};
\node[vert] (v00) at (-0.125,0.125) {$\overline{00}$};
\node[vert] (v01) at (-0.125,0.375) {$\overline{01}$};
\node[vert] (v10) at (-0.125,0.625) {$\overline{10}$};
\node[vert] (v11) at (-0.125,0.875) {$\overline{11}$};

\node[interleaved] at (h00 |- v00) {\fancylabel{0}{0}{0}{0}};
\node[interleaved] at (h00 |- v01) {\fancylabel{0}{0}{1}{0}};
\node[interleaved] at (h00 |- v10) {\fancylabel{1}{0}{0}{0}};
\node[interleaved] at (h00 |- v11) {\fancylabel{1}{0}{1}{0}};
\node[interleaved] at (h01 |- v00) {\fancylabel{0}{0}{0}{1}};
\node[interleaved] at (h01 |- v01) {\fancylabel{0}{0}{1}{1}};
\node[interleaved] at (h01 |- v10) {\fancylabel{1}{0}{0}{1}};
\node[interleaved] at (h01 |- v11) {\fancylabel{1}{0}{1}{1}};
\node[interleaved] at (h10 |- v00) {\fancylabel{0}{1}{0}{0}};
\node[interleaved] at (h10 |- v01) {\fancylabel{0}{1}{1}{0}};
\node[interleaved] at (h10 |- v10) {\fancylabel{1}{1}{0}{0}};
\node[interleaved] at (h10 |- v11) {\fancylabel{1}{1}{1}{0}};
\node[interleaved] at (h11 |- v00) {\fancylabel{0}{1}{0}{1}};
\node[interleaved] at (h11 |- v01) {\fancylabel{0}{1}{1}{1}};
\node[interleaved] at (h11 |- v10) {\fancylabel{1}{1}{0}{1}};
\node[interleaved] at (h11 |- v11) {\fancylabel{1}{1}{1}{1}};

\node[li] at (0.25 ,0.25) {$\ammcode{0}{0}$    $\Leftrightarrow$  \\$\fancylabel{0}{0}{\plh}{\plh}$    $\Leftrightarrow$    $Q_{\smallvector{0}{0},\smallvector{1}{1}}$};
\node[li] at (0.25 ,0.75) {$\ammcode{0}{1}$    $\Leftrightarrow$  \\$\fancylabel{1}{0}{\plh}{\plh}$    $\Leftrightarrow$    $Q_{\smallvector{0}{1},\smallvector{1}{1}}$};
\node[li] at (0.625,0.49) {$\ammcode{10}{}$ \\[1pt] $\Updownarrow   $ \\[5pt] $\fancylabel{\plh}{1}{\plh}{0}$ \\[5pt] $\Updownarrow   $ \\ $Q_{\smallvector{2}{0},\smallvector{2}{0}}$};
\node[li] at (0.875,0.49) {$\ammcode{11}{}$ \\[1pt] $\Updownarrow   $ \\[5pt] $\fancylabel{\plh}{1}{\plh}{1}$ \\[5pt] $\Updownarrow   $ \\ $Q_{\smallvector{3}{0},\smallvector{2}{0}}$};
\end{tikzpicture}
    \caption{
        Adapted from \cref{fig:before:square}:
        $2$-d illustration of the relation between AMM~\cite{bhatiaAMMAdaptiveMultilinear2022} location codes (in round brackets), Z-order location code matrix (over- and underlined numbers), and level-index notation ($Q_{\indexd,\level}$).
        The AMM location codes are separated by dimension, and the length of the binary string per dimension determines the refinement level $\level$.
        The Z index is obtained by interleaving the bits of the binary representation of each dimension's indexes (at the maximum resolution present);
        This is denoted by the red and blue binary numbers in the graphic's background.
        For larger rectangles, one can summarize several of the Z indices with \enquote{don't care} placeholders ($\diamond$), where the digits do not need to be considered to define the rectangle.
        The same can be obtained by interleaving the bits of the AMM location codes and filling in the missing places with $\diamond$ placeholders.
        Equivalently, the level-index notation of a rectangle is given by $Q_{\indexd,\level}$, with index \indexd{} and level \level{}.
        The level indicates how many bits should be considered in a particular dimension, and the index is the number extracted from the binary representation of the considered bits in that dimension.
        For every node in the omnitree, the labels of all ancestor nodes accumulated give the level of the $Q_{\indexd,\level}$ that the node is referring to, cf. \cref{fig:before:tree}.
    }
    \label{fig:morton}
\end{figure*}

Every node in an omnitree (\cref{def:omnitree}) corresponds to a well-defined rectangle $Q_{\indexd{},\level{}}$. The hierarchical level \level{} of a node in the tree can be obtained by accumulating all its ancestor's labels in a dimension-wise fashion; 
For instance, the four finest rectangles in \cref{fig:before:square} have one of two possible shapes, given by levels $\level = (1,1)$ and $\level = (2,0)$, which are obtained from adding the root's and the immediate parents' respective labels as seen in \cref{fig:before:tree}.

For a given rectangle $Q$, our refinement algorithm also refers to its AMM~\cite{bhatiaAMMAdaptiveMultilinear2022} location codes, which we abbreviate as $\stringify{Q}$.
The relation between AMM~\cite{bhatiaAMMAdaptiveMultilinear2022} location codes, Z order location code matrix (\enquote{Morton code}), and the level-index notation in this case is illustrated in \cref{fig:morton}.
When adding the information that a rectangle $Q$ has splitting label $\vec{b}$, we create \textit{extended location codes} $\lambdify{Q}$ by appending a $(\lambda)$ character to $\stringifyInDimension{j}{Q}$ in each of the split dimensions $(j:b_j=1)$.
For illustrative purposes, we aggregate the extended location codes of all nodes in an omnitree into a combined \textit{location stack} (\crefrange{fig:refinement_sweep:initial_location_stack}{fig:refinement_sweep:refined_location_stack}).

For complexity analysis, we use the following notations:
$d$ (as before) is the dimensionality.
$N$ denotes the number of most-refined rectangles in the domain $\Omega$ or equivalently, the number of leaf nodes in the omnitree.
$n_m$ denotes the number of (one-dimensional) refinements requested to obtain the refined tree.{}
$h$ denotes the mean tree depth of the omnitree as observed by the leaf nodes.
Assuming that the tree is refined at the locus of a subset $\Omega^* \subset \Omega$ with positive measure $0 < d^* \leq d$, $N$ is dominated by an exponential number of deepest leafs and $h(N) \in \Theta(\frac{1}{d^*}\log(N))$.
For instance, $d^*=2$ for a surface.
In the exceptional case that $d^*=0$ (a point particle), the tree may be totally skewed with only a constant number of leafs per level and $h(N) \in \Theta(N)$.

Top-down construction of an omnitree begins with the simplest omnitree, consisting of a singleton root--leaf node spanning all of $\Omega$ and labeled with $d$ zeros.
\begin{figure*}
    \centering
    \begin{subfigure}[t]{0.32\textwidth}
        \centering
        \hspace*{-2em}\includestandalone[width=\textwidth]{gfx/tree_initial_label}
        \caption{Initial markers.}
        \label{fig:refinement_sweep:initial}
    \end{subfigure} 
    \begin{subfigure}[t]{0.32\textwidth}
        \centering
        \hspace*{-2em}\includestandalone[width=\textwidth]{gfx/tree_swept_up}
        \caption{After sweeping.}
        \label{fig:refinement_sweep:up}
    \end{subfigure}
    \begin{subfigure}[t]{0.34\textwidth}
        \centering
        \hspace*{-2em}\includestandalone[width=\textwidth]{gfx/tree_refined}
        \caption{Resulting refined tree.}
        \label{fig:refinement_sweep:refined}
    \end{subfigure}\\
    \begin{subfigure}[t]{0.32\textwidth}
        \centering
        \definecolor{color_0}{RGB}{8,92,248}
\definecolor{color_1}{RGB}{103,175,29}
\definecolor{color_2}{RGB}{244,204,31}
\definecolor{color_3}{RGB}{252,105,156}
\begin{tikzpicture}[
       every node/.style={align=left,anchor=west,
       text height=2ex,minimum width=2ex,
       inner sep=0.2ex,
       fill opacity=0.4, text opacity=1}]
    \matrix [
       draw=none, matrix of nodes] (n)
    {    \texttt{\scalebox{0.85}{$\lambda$}}&\texttt{}\\
    \texttt{0}&\texttt{\scalebox{0.85}{$\lambda$}}\\
    |[fill=color_0]| \texttt{0}&|[fill=color_0]| \texttt{0}\\
    |[fill=color_1]| \texttt{0}&|[fill=color_1]| \texttt{1}\\
    \texttt{1\scalebox{0.85}{$\lambda$}}&\texttt{}\\
    |[fill=color_2]| \texttt{10}&|[fill=color_2]|\texttt{+}\\
    |[fill=color_3]| \texttt{11}&|[fill=color_3]|\texttt{+}\\
};
\draw (n-1-1.north west) -- (n-7-1.south west)(n-1-2.north west) -- (n-7-2.south west);
\end{tikzpicture}
        \caption{Location stack for (a).}
        \label{fig:refinement_sweep:initial_location_stack}
    \end{subfigure} 
    \begin{subfigure}[t]{0.32\textwidth}
        \centering
        \definecolor{color_0}{RGB}{8,92,248}
\definecolor{color_1}{RGB}{103,175,29}
\definecolor{color_2}{RGB}{244,204,31}
\definecolor{color_3}{RGB}{252,105,156}
\begin{tikzpicture}[
       every node/.style={align=left,anchor=west,
       text height=2ex,minimum width=2ex,
       inner sep=0.2ex,
       fill opacity=0.4, text opacity=1}]
    \matrix [
       draw=none, matrix of nodes] (n)
    {    \texttt{\scalebox{0.85}{$\lambda$}}&\texttt{+}\\
    \texttt{0}&\texttt{\scalebox{0.85}{$\lambda$-}}\\
    |[fill=color_0]| \texttt{0}&|[fill=color_0]| \texttt{0}\\
    |[fill=color_1]| \texttt{0}&|[fill=color_1]| \texttt{1}\\
    \texttt{1\scalebox{0.85}{$\lambda$}}&\texttt{}\\
    |[fill=color_2]| \texttt{10}&\texttt{}\\
    |[fill=color_3]| \texttt{11}&\texttt{}\\
};
\draw (n-1-1.north west) -- (n-7-1.south west)(n-1-2.north west) -- (n-7-2.south west);
\end{tikzpicture}
        \caption{Location stack for (b).}
        \label{fig:refinement_sweep:up_location_stack}
    \end{subfigure}
    \begin{subfigure}[t]{0.32\textwidth}
        \centering
        \definecolor{color_0}{RGB}{8,92,248}
\definecolor{color_1}{RGB}{59,158,74}
\definecolor{color_2}{RGB}{153,187,24}
\definecolor{color_3}{RGB}{244,204,31}
\definecolor{color_4}{RGB}{254,144,123}
\definecolor{color_5}{RGB}{246,66,152}
\definecolor{previous_color_0}{RGB}{8,92,248}
\definecolor{previous_color_1}{RGB}{103,175,29}
\definecolor{previous_color_2}{RGB}{244,204,31}
\definecolor{previous_color_3}{RGB}{252,105,156}
\begin{tikzpicture}[
       every node/.style={align=left,anchor=west,
       text height=2ex,minimum width=2ex,
       inner sep=0.2ex,
       fill opacity=0.4, text opacity=1}]
    \matrix [
       draw=none, matrix of nodes] (n)
    {    \texttt{\scalebox{0.85}{$\lambda$}}&\texttt{\scalebox{0.85}{$\lambda$}}\\
    |[fill=previous_color_0]| \texttt{0}&|[fill=previous_color_0]| \texttt{0}\\
    \texttt{1\scalebox{0.85}{$\lambda$}}&\texttt{0}\\
    |[fill=previous_color_2]| \texttt{10}&|[fill=previous_color_2]| \texttt{0}\\
    |[fill=previous_color_3]| \texttt{11}&|[fill=previous_color_3]| \texttt{0}\\
    |[fill=previous_color_1]| \texttt{0}&|[fill=previous_color_1]| \texttt{1}\\
    \texttt{1\scalebox{0.85}{$\lambda$}}&\texttt{1}\\
    |[fill=previous_color_2]| \texttt{10}&|[fill=previous_color_2]| \texttt{1}\\
    |[fill=previous_color_3]| \texttt{11}&|[fill=previous_color_3]| \texttt{1}\\
};
\draw (n-1-1.north west) -- (n-9-1.south west)(n-1-2.north west) -- (n-9-2.south west);
\end{tikzpicture}
        \caption{Location stack for (c).}
        \label{fig:refinement_sweep:refined_location_stack}
    \end{subfigure}
    \caption{
        Markers for the refinement algorithm applied to the tree in \cref{fig:before:tree}:
        The two rightmost leaf nodes are marked for refinement in the second dimension. 
        After the upwards sweep, this refinement is placed at the root node of the tree, and a negative refinement is marked for the left parent node.
        For this tree, there is no change after the downward sweep, since the refinements can each be realized at the nodes that they are now placed at.
        The negative refinement means that the left parent node vanishes and its children are adopted by the root node in the refined tree.
        Conversely, the right parent node is now duplicated and the duplicates need to be interleaved with the left leaf nodes to comply with the Z order.
        (The nodes in \cref{fig:refinement_sweep:refined} use the \enquote{old} colors to allow for easier comparison with the initial tree; the actual color ordering in the new tree can be seen in \cref{fig:after:tree}.)
        This example illustrates the importance of step \ref{alg:refinement:reordering} in the algorithm.
    }
    \label{fig:refinement_sweep}
\end{figure*}

\begin{algorithm}[htbp!]
  \caption{Construction algorithm. \\
    \textbf{Input:} Rectangle $q$ identified by $\stringify{q}$ whose new subtree $V$ should be constructed, source subtree $T$ with root $t$ whose rectangle covers $q$\\
    \textbf{Output:} Subtree $V$ of the target tree $T$ (the result of applying refinement markers to $T$)
    }
  \begin{algorithmic}[1]
            \Procedure{ConstructNewTree}{$q, T$}
    \If{$\vec{b}(t) = \vec{0}$} \Comment{if $t$ is a leaf in the source tree}
    \State $V := $ tree (root rectangle $q$), new descendants according to $\vec{m}(t)$
    \State \Return $V$
    \Else
    \State $S = $\Call{SearchDescendant}{$q, T$}
    \State $s = \text{root}(S)$
    \State $v = $ new node
    \State $\vec{b}(v)= \vec{b}(s) + \vec{m}(s)$
    \State $\sigma(v) \equiv \{j \in \mathcal{D}\colon b_j(v)=1\}$
    \State $\lambdify{q} \coloneqq \stringify{q}$
    \ForAll{dimensions $j \in \sigma(v)$}  \Comment{Mutate $\lambdify{q}$}
      \State $\lambdifyInDimension{j}{v} \coloneqq \stringifyInDimension{j}{v} +$ \enquote{$\lambda$}   \Comment{String concatenation}
    \EndFor
    \If{not $\vec{b}(v) = \vec{0}$} \Comment{if $v$ is not a leaf in the target tree}
    \ForAll{child indices $e \in \kappa(v)$}
      \State $w_e =$ rectangle such that $\stringify{w_e} = \stringify{q} + e$  \Comment{subrectangle of $q$}
      \State $W_e = $ \Call{ConstructNewTree}{$w_e$, $S$}
      \State $V[e] \coloneqq W_e$  \Comment{Attach $W_e$ as a subtree}
    \EndFor
    \EndIf
    \State \Return{$V$}
    \EndIf
    \EndProcedure
  \end{algorithmic}
  \label{alg:refinement:construction}
\end{algorithm}
Then, the refinement algorithm consists of the following steps, which can be iterated:
\begin{enumerate}
    \item \emph{Attach refinement markers to nodes}: 
    A refinement marker is a vector $\vec{m} \in \positived$, whose components $m_j$ indicate how many refinement levels should be added to this node in dimension $j$.
    Refinement markers are applied to many nodes at once, regardless of whether they are leaf nodes; $n_m = \sum \abs{m}$.

    Complexity: $\complexity{n_m}$.
    
    \item \emph{Sweep refinement markers bottom-up}:
    Starting at the bottom of the tree, one checks for each refinement in a marker whether it can be moved to the parent node.
    This is the case if all siblings either have common refinement markers or a \enquote{1} refinement at the relevant dimension.
    In the latter case, the lifting of this refinement will be indicated by placing a negative marker for the respective sibling.

    Complexity: $\complexity{n_m \log{n_m} + \min(N, h \cdot n_m})$.
    
    \item \emph{Sweep refinement markers top-down}:
    At the end of the last step, markers will have moved upwards as far as possible, which may be too far up;
    potentially, there now are refinement markers at nodes that are already refined in the respective dimension.
    For this reason, starting at the top of the tree, one checks whether the markers' refinement can be resolved at the current node.
    If not, the refinement(s) that cannot be resolved are pushed down to the children of the node.

    Complexity: $\complexity{n_m \log{n_m} + \min(N, h \cdot n_m)}$.
    
    \item \emph{Construct the refined tree by reordering parent nodes and expanding leafs}:
    \label{alg:refinement:reordering}
    The preceding steps in the refinement algorithm annotate the source tree with appropriately placed refinement markers.
    The final step constructs the target tree top-down, applying refinements along the way.
    Starting with the full source tree as $T$ and its root node as $t$ (with $\stringify{t}$ empty), one can apply the recursive procedure \textsc{ConstructNewTree}, in \cref{alg:refinement:construction}.
    If $t$ is a leaf node, the full tree is simply obtained by expanding the refinement marker $\vec{m}(t)$ to create descendants of $t$.
    At nonleaf nodes, there can be significant structural changes.
    The new splitting label and thus number of children may differ from the original.
    Positive markers double the branching factor, requiring grandchild nodes to be adopted, while negative markers halve the branching factor, sometimes leading to redundant, unsplit nodes that must be culled.
    (\Cref{fig:refinement_sweep} illustrates the simplest two-dimensional case where a non-leaf node becomes redundant and vanishes after refinement---it is integrated into its own parent node.
    See also \cref{fig:2d_gradchild_split} for the respective discretizations of the unit square.)
    In extreme cases, the source tree may not have any nodes in common with the target tree besides the root and leafs.
    Therefore, we must map each constructed node into the source tree to find the relevant splitting label (\cref{alg:refinement:searching}).{}
    Accordingly, it is necessary that \textsc{ConstructNewTree} calls \textsc{SearchDescendant} in \cref{alg:refinement:searching} to identify the source tree node $s$ that either already corresponds to the same rectangle $q$, or that covers $q$ (it will be split to match $q$).
    The splitting label of the newly constructed node $v$ can be obtained by summing $s$' label and the marker attached to $s$.
    In the location stack in \cref{fig:refinement_sweep:up_location_stack}, the addition is interpreted as replacing $(+)$ markers by $(\lambda)$ labels and annihilating $(\lambda,-)$ pairs.
    To form $\lambdify{v}$, a label $(\lambda)$ is appended to $\stringify{v}$ in each split dimension $j \in \sigma'(v)$.
    This completes construction of node $v$ in the target tree.
    The location codes $\stringify{w_e}$ of its child nodes $w_e$ are obtained from $\lambdify{v}$ by substituting trailing $(\lambda)$ labels with $0$ or $1$.
    The tree is then built up recursively, by calling \textsc{ConstructNewTree} for every child rectangle $w_e$, referring to the source subtree originating at node $s$.
    

    


    


    Complexity: $\complexity{N + \min(N, n_m \cdot h \cdot 2^d)}$, for copy and reordering, respectively.

\end{enumerate}
        
        
        
        

\begin{algorithm}[htb!]
    \caption{Search algorithm. \\
    Notations: $\abs{\cdot}$ for string length, $\text{string}[\text{mask}]$ for binary sub-string selection.\\
    \textbf{Input:} Rectangle $q$ identified by $\stringify{q}$, tree $T$ with root $t$ whose rectangle $r$ covers $q$ \\
    \textbf{Output:} Smallest subtree $\subset T$ whose root rectangle covers $q$
    }
    \begin{algorithmic}[1]
    \Procedure{SearchDescendant}{$q$, $T$}
    
    \If{$\stringify{r} = \stringify{q}$}
    \State \Return T
    \EndIf
    \If{$\lvert \stringifyInDimension{j}{q} \rvert \geq \lvert \lambdifyInDimension{j}{r} \rvert$ $\forall j \in \mathcal{D}$}  \Comment{$q$ is a proper subrectangle of $r$}
    \State $e := $ child index $\big(\stringify{q}\big)\big[\lambdify{r}=\lambda\big] \in \kappa(t)$
    \State \Return \textsc{SearchDescendant}($q$, subtree $T[e]$)
    \Else
    \State \Return T \Comment{Since any subtree would only partially cover $q$}
    \EndIf
    \EndProcedure
    \end{algorithmic}
    \label{alg:refinement:searching}
\end{algorithm}

This algorithm is guaranteed to produce a new valid omnitree.
However, there are combinations of inputs that cause the tree to become un-normalized (losing property 3 in \cref{def:normalized_omnitree}).
This can be the case when multiple refinements in a dimension are applied to a node at once ($\max{\abs{m}} > 1$).
Our experiments in \cref{sec:results} only perform either refinement in all dimensions ($m = \vec{1}$) in the case of octrees, or in one dimension ($\sum{m} = 1, \max(m) = 1, \min(m) = 0$) in the case of omnitrees, on a single leaf node.
Therefore, all presented results are applicable to normalized omnitrees.{}

Our Python implementation~\cite{bleifreiFreifrauvonbleifreiDyAda2025} provides extensive tests and visualizations in addition to these procedures.

\subsection{Storage Costs and Convergence Rates}\label{subsec:storage_convergence}

The labeling scheme of the omnitrees allows to store them uniquely as a compact binary string representation, see for example \cref{fig:before:desc,fig:after:desc}.
It can be obtained by concatenating the labels of the tree in a preorder depth-first manner (as indicated by the dashed lines in \cref{fig:before:tree,fig:after:tree}).
This binary representation is conceptually similar to the DF-expression~\cite{kawaguchiMethodBinaryPictureRepresentation1980} while needing only two symbols ($1$ and $0$), since the function data is stored separately, outside the tree storage data structure.
The tree's storage cost is then given by $d$ bits times the number of tree nodes.



When considering only octrees---a special case of omnitrees---one can further compress the binary string, as the labels are going to be either all-one or all-zero, and they can be stored in as many bits as there are tree nodes.
From this fact, we can directly derive the added storage cost between omnitree and octree representations:
If the ideal discretization is given by an octree, then the tree will take $d$ times as much storage.
At the same time, the function values vector $\hat{g}$ will take up the same storage, namely the number of bits in the stored data type times the number of leaf nodes $N$.



This paper defines the (dimension-independent) convergence rate as
\begin{equation}
    r = \frac{\log\left(\frac{e_1}{e_2}\right)}{\log\left(\frac{N_2}{N_1}\right)}
\end{equation}
with $N_1$ and $N_2$ the number of rectangles in two approximations of the same problem, and errors $e_1$ and $e_2$ the respective application-dependent errors.
This definition allows us to derive the omnitree's maximum advantage for problems that exhibit extreme local anisotropy, for example when (after a given resolution) further refinement is only beneficial in a single dimension per rectangle.
Picture a cube with a real-valued function and high frequencies in that dimension, and constant values in all other dimensions.
In this case, the octree is going to create $2^d$ sub-rectangles where only two would have been necessary, creating $2^{d-1}$ \enquote{too many} leaf nodes, and enlarging $\hat{g}$ by the same factor.
However, this does not only affect the current refinement level;
To double the resolution in the dimension of interest again, there are now $2^{d-1}$ \enquote{too many} leaf nodes that need refining.
This is a factor that compounds over the different levels of resolution as long as refinement is only beneficial in a single dimension.
Accordingly, a very anisotropic problem can see a degradation of up to a factor of $d$ in the convergence rate if octrees are used in place of omnitrees:
\begin{equation}
    e^\text{oct}_1 \leq e^\text{omni}_1, e^\text{oct}_2 \leq e^\text{omni}_2 \Rightarrow 1 \geq \frac{r^\text{oct}}{r^\text{omni}} \geq \frac{\log(2)}{\log(2^d)} = \frac{1}{d}
\end{equation}
Conversely, omnitrees can increase the convergence rate by up to a factor of $d$;
This estimate is independent of the order of convergence of the method itself.


For most computational problems, the data vector $\hat{g}$ will be the dominating factor in storage compared to the tree representation.
As a coarse approximation, an omnitree always has fewer parent nodes than leaf nodes, so the length of the binary string is bounded by $2N \cdot d$ bits.
For a typical simulation problem, one may deal with $d=3$ dimensions and 32-bit floating point values in $\hat{g}$, such that the tree data will take up less than one fifth of the data vector $\hat{g}$ (less than \num{6}$N$ vs. $32N$ \si{\bit}).
As soon as there is any anisotropy occurring at the resolved scales, it is very likely that the compounding savings in the omnitree will lead to lower storage for the same accuracy.

In fact, the experiments in \cref{sec:eval} describe an extreme case where $d=3$ or $d=4$ and the data type in $\hat{g}$ is a 1-bit binary (as small as possible).
Accordingly, in the most isotropic scenarios, the omnitree will need less than $\frac{7}{3}$ times the storage for tree and data storage, compared to the (single-digit-labeled) octree.
In all other scenarios, the omnitree advantage will become more and more visible for higher resolutions, by a factor of up to $d$ per scale, and quickly outweigh the constant storage overhead.
(In \cref{sec:results}, we are going to observe that the latter is the more realistic case for our $3$-d and $4$-d objects).



In summary, while there may be moderate constant-factor storage costs to using omnitrees in place of octrees for isotropic problems, significant benefits can be expected for anisotropic problems, and especially for high dimensionalities.

\section{Evaluation Method on Shapes}\label{sec:eval}

For evaluation of omnitrees compared to octrees, this paper tries to utilize the simplest possible case: 
$3$-d shape representation, and $4$-d shapes through time-dependent rotation.
This is suitable, as three dimensions are the highest dimensionality for which octrees are commonly used, and the shape surfaces can be visualized well.

We use the dataset \thingitenk{}~\cite{zhouThingi10KDataset100002016}, a dataset consisting of \num{10000} $3$-d printable objects published on the Thingiverse platform.
Of the \num{10000} \enquote{thingies}, we select the \num{4166} objects that have up to \num{10000} vertices, and that are watertight, not self-intersecting, and solid.
As a preprocessing step, we ensure that the objects are placed within the unit cube:
We translate and uniformly scale each object such that the minimum and maximum of the largest axis-aligned dimension reaches exactly the interval $[0,1]$, and the other dimensions are centered around \num{0.5}.
For a select set of three-dimensional objects, we design a time-dependent (four-dimensional) problem by rotating the object around a diagonal axis ($(1, 1, 1)^T$) and re-scaling like above for every given time.
These three- to four-dimensional objects are then discretized onto octree and omnitree grids.
The $L_1$ (mass) error is evaluated as $\norm{g' - g^*}_{1}$, where the ground truth discriminator $g^*(x) : \Omega \rightarrow \{0,1\}$ returns \num{1} precisely when $x$ is in the object, and the approximation $g'$ is piecewise constant over rectangles.
Our experiment samples $g^*(x)$ pointwise and sets $g'(x)$ as the majority value within each rectangle.

To construct an efficient omnitree, the top-down algorithm in \cref{sec:refinement} needs an estimate of the utility of refining any set of dimensions $\mathcal{I} \subset \mathcal{D}$ in any rectangle $Q$.
Our experimental methodology uses variance-based sensitivity analysis~\cite{saltelliGlobalSensitivityAnalysis2008} to estimate this utility.
In particular, for any multivariable function $f: Q \to \mathbb{R}$, the Sobol' variance decomposition,
\begin{equation}
  \begin{aligned}
  \label{eq:sobol-variance}
    \sum_{\mathcal{I} \subset \mathcal{D}} \variance[f_\mathcal{I}] &= \variance[f] \\
    \variance[f_\mathcal{I}] + \sum_{\mathcal{J} \subsetneq \mathcal{I}} \variance[f_\mathcal{J}] &=
    \variance[\expectation[f(x) \mid x_i,\, i\in\mathcal{I}]], \quad \variance[f_\emptyset] = 0
  \end{aligned}
\end{equation}
is a Pythagorean formula for the variance $\variance[f]$ that reveals orthogonal components $\variance[f_\mathcal{I}]$ along functions of different subsets of dimensions.
Each component may be computed by evaluating the conditional expectation $\expectation[f(x) \mid x_i,\, i\in\mathcal{I}]$ and excluding lower order components.
When normalized by the total variance, these components produce the \emph{Sobol' sensitivity indices},
$S_\mathcal{I} \coloneqq \variance[f_\mathcal{I}] / \variance[f]$.


Taking $f=g^*$, the first order variances
$\variance[g_1],\, \variance[g_2],\, \variance[g_3],\, (\variance[g_4])$
quantify the individual importance of resolving a given dimension $i \in \mathcal{D} = \{1,2,3,4\}$ in rectangle $Q$.
To prioritize dimension-wise refinements across the entire omnitree, the local first order variances are scaled by rectangle volume, $\variance[g_i]\cdot|Q|$.
The refinement criterion for the octree is simply $\variance[g^*]\cdot|Q|$, as refinements are applied to all dimensions or none.
For either type of tree, rectangles with no variance, for example, those entirely inside or outside the object, receive a priority of \num{0}.

For practical use, the experiments approximate the sensitivity indices through the $\delta$ moment-independent analysis method developed by Borgonovo~\cite{borgonovoNewUncertaintyImportance2007} applied to samples obtained by Saltelli's method~\cite{saltelliMakingBestUse2002} (both as implemented in SALib~\cite{HermanSALib2017,IwanagaSALib2022}).
Saltelli sampling is parameterized on the number of samples $n_s$, which needs to be a power of two, resulting in a total of $n_s \cdot (2d + 2)$ coordinates at which $g^*$ has to be evaluated.
$g^*(x)$ is evaluated pointwise (via the trimesh~\cite{trimesh} library's \texttt{contains} function).
We use $n_s = \num{512}$, i.e., \num{4096} sample points per cuboid in three dimensions and \num{8192} sample points per hyper-rectangle in four dimensions.
During refinement, these values are stored in a priority queue, with only the highest-priority refinements being realized, until a certain number of rectangles in the discretization is surpassed; we choose powers of two starting from \num{16} for this purpose.

Once the target number of rectangles is reached, one can calculate the corresponding data vector $\hat{g}$ (\cref{eq:discrete-function}), again by a Monte Carlo method.
For each rectangle in the discretization, we take $n_g = \num{4096}$ uniformly sampled random points in $3$-d and $n_g = \num{8192}$ in $4$-d.
We assign either $0$ or $1$ for the corresponding entry in $\hat{g}$, based on the majority of $g^*$.
Finally, we can approximate the $L_1$ error between $g'$ and the ground truth $g^*$ by randomly sampling from the whole unit cube domain $\Omega$:
\begin{equation}\label{eq:monte_carlo_l1}
\norm{ g' - g^*}_{1} \approx \sum_{i=1}^{N_e} \abs{g'(\mathbf{x}_i) - g^*(\mathbf{x}_i)}
\quad \text{with} \quad \mathbf{x}_i \sim \text{Uniform}(\Omega).
\end{equation}
Another option would be to use the relative error, through an additional division by $\norm{g^*}$; 
However, both error measures skew the error depending on the volume of the object.
The absolute error results in rather small initial errors for small objects, whereas the relative error will return comparably large values for small objects.
To keep the evaluation as simple as possible, we choose the absolute Monte Calro $L_1$ error.
For the purposes of evaluating the error, the experiments use $n_e= 2^{18} = \num{262144}$ samples in $3$-d and $2^{24} = \num{16777216}$ in $4$-d.

\section{How Well Can Shapes Be Approximated?---Octrees vs. Omnitrees}\label{sec:results}

This section evaluates the aggregate $L_1$ error statistics on the \num{4166} three-dimensional objects when approximated as octrees and omnitrees, as described in the last chapter.
In addition, we present some interesting insights on the memory usage and information density of the octree and omnitree representations.
For a select subset of $3$-d and $4$-d objects, we provide visualizations and a closer look and interpretation of their error characteristics at different resolutions.
Finally, we apply the refinement algorithms to a real-world usage example, the F25 aircraft model.

\subsection{\num{4166} Three-Dimensional Objects in Three Dimensions}

\begin{figure}
    \centering
    \def\axisdefaultwidth{10cm}
\def\axisdefaultheight{6cm}
\pgfplotsset{
	treebox/.style={
		boxplot prepared={
			lower whisker=1e-10, 
			every box/.style={draw=black},
			every median/.style={draw=black},
            box extend=0.5,
		},
		boxplot prepared,
		boxplot,
		fill,
		mark size=0.8pt,
		opacity=0.8,
	},
	meanline/.style={
		very thick,
		table/y expr=\thisrow{mean},
		table/x expr=\coordindex,
	},
	contourline/.style={
		meanline,
		ultra thick,
		black,
	},
}
\begin{tikzpicture}
	\pgfplotstableread[col sep=comma]{l1_error_octree_s512.csv}\aggregateomnioct
	\pgfplotstableread[col sep=comma]{l1_error_omnitree_1_s512.csv}\aggregateomnione
	\pgfplotstableread[col sep=comma]{l1_errors_octree_s512.csv}\csvdataoct
	\pgfplotstableread[col sep=comma]{l1_errors_omnitree_1_s512.csv}\csvdataone
	\begin{semilogyaxis}[
		boxplot/draw direction = y,
		errorplot,
		ymax = 0.3,
		ymin = 0.0002,
		xmin = -0.5,
		xmax = 9.5,
		enlarge y limits,
		ymajorgrids,
		xtick = {0,...,9},
		xticklabel style = {align=center, font=\small, rotate=60},
		xticklabels = {16, 32, 64, 128, 256, 512, 1024, 2048, 4096, 8192},
        legend pos=south west,
	]
		\addplot[octree, meanline] table {\aggregateomnioct};
		\addlegendentry{Octree}
		\addplot[omnione, meanline] table {\aggregateomnione};
		\addlegendentry{Omnitree}
		\foreach \n in {0,...,9} {
			\addplot[
				octree,
				treebox,
				xshift=-2pt,
				boxplot prepared={
					draw position=\n,
				},
				boxplot,
			] table[y index=\n]\csvdataoct;
			\addplot[
				omnione,
				treebox,
				xshift=2pt,
				boxplot prepared={
					draw position=\n,
				},
				boxplot, 
			] table[y index=\n]\csvdataone;
		}
		\addplot[octree, contourline] table {\aggregateomnioct};
		\addplot[octree, meanline] table {\aggregateomnioct};
		\addplot[omnione, contourline] table {\aggregateomnione};
		\addplot[omnione, meanline] table {\aggregateomnione};
	\end{semilogyaxis}
\end{tikzpicture}
    \caption{
        $L_1$ errors for the \thingitenk{} data set over the number of cuboids in the discretization for octree and omnitree refinement.
        Experimental parameters are as described in \cref{sec:eval}.
        The mean values are indicated by the thick line plot.
        For the maximum number of cuboids tested (\num{8192}), the difference in the mean error between octree and omnitree amounts to a factor of \num[round-mode=places,round-precision=1]{5.65681126}, 
        The boxplots in the background give more distributional information:
        The median values are significantly lower than the mean values, and the difference in the median error is \num[round-mode=places,round-precision=1]{10.135638298}${\times}$.
        The fact that the q25 and q75 percentiles appear almost equally large on the logarithmic scale actually means that the density of values is highest below the median. 
        The lower whiskers are cut off, as they would actually extend to 0, which cannot be represented on the log scale (there are cubes and cuboids in the dataset that can be perfectly resolved, even at low resolutions).
    }
    \label{fig:results:errors10k}
\end{figure}

\Cref{fig:results:errors10k} shows the aggregate results in the Monte-Carlo $L_1$ error (\cref{eq:monte_carlo_l1}).{}
The distribution of errors among the $3$-d objects for every resolution $N$ is very skewed towards low values.
Accordingly, the mean values are dominated by high error values, and we mostly take median values for our further analysis for more representative results.
Towards the right of the graph, one can observe a convergence rate of the mean error of \num[round-mode=places,round-precision=2]{0.388425925} for the octree, and \num[round-mode=places,round-precision=2]{0.592861019} for the omnitree implementation of the shape approximation.
The convergence rate in the median error for the octree amounts to \num[round-mode=places,round-precision=2]{0.563989956}, while one can observe \num[round-mode=places,round-precision=2]{0.760560711} for the omnitree.
Conclusively, in both measures, the convergence rate is increased by \num{0.20} when using omnitree instead of octree discretization.
In summary, most of the $3$-d objects can be resolved rather well at the resolutions considered in the experiments, and the mean error is dominated by the relatively few that cannot;
However, the omnitree measurably increases the order of convergence over the octree in all statistical measures for this data set.
As expected, the convergence rates' statistical difference, as well as the individual difference per object, lie within the range of a factor of \num{1} (no anisotropy) and $d = 3$ (extreme anisotropy).

\begin{figure}
    \centering
    \begin{subfigure}[t]{0.47\textwidth}
    \def\axisdefaultwidth{10cm}
\def\axisdefaultheight{7cm}
\pgfplotsset{
	scatter/.style={
		only marks,
		mark size=0.1pt,
	},
	medianmarkers/.style={
		mark size=3pt,
		table/x expr=\thisrow{median_total_storage_size},
		table/y expr=\thisrow{median},
	},
	contourmarkers/.style={
		medianmarkers,
		ultra thick,
		black,
	},
}
\begin{tikzpicture}
	\pgfplotstableread[col sep=comma]{../data/l1_error_octree_s512.csv}\aggregateomnioct
	\pgfplotstableread[col sep=comma]{../data/l1_error_omnitree_1_s512.csv}\aggregateomnione
	\begin{axis}[
		xmode=log,
		ymode=log,
		ymin = 5e-4,
		ymax = 0.1,
		grid=both,
		xticklabel style = {align=center, font=\small},
		ylabel = {Median $L_1$ error (Monte Carlo)},
		xlabel = {\shortstack{Median storage size (tree and data vector)}},
		x unit = \si{bit},
        legend pos=south west,
	]
		\addplot[octree, medianmarkers] table {\aggregateomnioct};
		\addlegendentry{Octree}
		\addplot[omnione, medianmarkers] table {\aggregateomnione};
		\addlegendentry{Omnitree}
		\addplot[octree, medianmarkers] table {\aggregateomnioct} coordinate (lastoct) coordinate[pos=0] (firstoct);
		\addplot[omnione, medianmarkers] table {\aggregateomnione} coordinate (lastomni) coordinate[pos=0] (firstomni);
		\node (ellipsemid) at (2000,0.013) {};
		\node (ellipselabel) at (20000,0.02) {};
	\end{axis}
    \draw (ellipsemid) ellipse [x radius=1.5cm, y radius=0.2cm];
	\node[align=center,thick] at (ellipselabel) {\textbf{5x less memory}\\for similar error};
	\node [octree_t,thingilabel_left] at (firstoct){16};
	\node [octree_t,thingilabel] at (lastoct){8192};
	\node [omnione_t,thingilabel_left] at (firstomni){16};
	\node [omnione_t,thingilabel_left] at (lastomni){8192};
\end{tikzpicture}
    \caption{
        Median $L_1$ error per discretization over the median memory footprint (tree and data vector combined).
    }
    \label{fig:results:memory}
    \end{subfigure}\hfill
    \begin{subfigure}[t]{0.47\textwidth}
        \def\axisdefaultwidth{10cm}
\def\axisdefaultheight{7cm}
\pgfplotsset{
	medianline/.style={
		very thick,
		table/x expr=\thisrow{median_shannon_information_function}, 
		table/y expr=\thisrow{median},
	},
}
\begin{tikzpicture}
	\pgfplotstableread[col sep=comma]{../data/l1_error_octree_s512.csv}\aggregateomnioct
	\pgfplotstableread[col sep=comma]{../data/l1_error_omnitree_1_s512.csv}\aggregateomnione
	\begin{axis}[
		ymode=log,
		grid=both,
		xmin = 0,
        xmax = 1,
		ymin = 5e-4,
		ymax = 0.1,
		ylabel = {Median $L_1$ error (Monte Carlo)},
        xlabel = {Median Information Density},
        legend pos=south west,
	]
		\addplot[octree, medianline] table {\aggregateomnioct} coordinate (lastoct) coordinate[pos=0] (firstoct);
		\addlegendentry{Octree}
		\addplot[omnione, medianline] table {\aggregateomnione} coordinate (lastomni) coordinate[pos=0] (firstomni);
		\addlegendentry{Omnitree}
	\end{axis}
	\node [octree_t,thingilabel_left] at (firstoct){16};
	\node [octree_t,thingilabel_left] at (lastoct){8192};
	\node [omnione_t,thingilabel_up] at (firstomni){16};
	\node [omnione_t,thingilabel_left] at (lastomni){8192};
\end{tikzpicture}
        \caption{
            Median $L_1$ error over the median information density, \cref{eq:shannon_density}.
        }
        \label{fig:results:information}
    \end{subfigure}
    \caption{
        Memory footprint (left) and information convergence (right) for the \thingitenk{} data set.
        Values are taken as median per discretization; 
        The considered discretizations are the same as in \cref{fig:results:errors10k}, with 16 to \num{8192} cubes or cuboids as indicated by the numbers.
    }
\end{figure}

\Cref{subsec:storage_convergence} described the storage overhead for omnitrees, but also estimated that the overhead would quickly be amortized by the reduced storage costs for the function data and the compounding increased convergence.
From \cref{fig:results:memory}, one can infer that this indeed holds true:
With omnitrees, one can obtain lower errors at the same storage size (when reading \cref{fig:results:memory} vertically) or spend less storage to get the same level of error (when reading horizontally).
The tree storage for the octree was one bit per tree node, and the omnitree uses three bits per tree node.
Note that the data sizes used here are an extreme example where the tree storage is always larger than the data vector's storage;
For applications in simulations where the data consists of \SIrange{16}{64}{\bit} floats, storage results are going to favor the omnitree even more.

\subsection{Convergence is High Information Density}

Our experiments approximate a binary-valued function $g$, indicating which areas of the domain are inside the object ($1$ denoting that a point is in the interior).  
The fact that $g$ is binary allows us to view the results from another angle:
We can analyze the information density in the data vector $\hat{g}$ by considering the number of ones and zeros stored in it. The Shannon information density of a bitstring is given by

\begin{equation}\label{eq:shannon_density}
H = -p_0 \log_2 p_0 - p_1 \log_2 p_1 ,
\end{equation}
where $p_0$ and $p_1$ are the proportion of zeros and ones, respectively.
The information density is maximized when both are equally likely ($p_0 = p_1 = \frac{1}{2}$).
\Cref{fig:results:information} shows the median $L_1$ error plotted over the median information density.
\footnote{We choose median values, since \cref{fig:results:errors10k} suggests that it may be the most expressive measure, but the results are qualitatively similar in the mean values.}
One can see a clear connection between the convergence of the $L_1$ error towards 0 and the convergence of $H$ towards $1$, meaning that the information density is maximized as the errors get lower and lower.
This can be explained as follows:
If the object surface is embedded into the domain without any particular orientation or structure, then, on average, we can expect that half of the newly constructed cuboids will lie mostly within the object, and the other half will be mostly outside the object.
This is more likely the higher the variance, and thus aligns with our refinement criteria.
As the refinement level increases in the regions of interest, the leaf nodes dominate the overall function vector, and they are exactly the nodes with this statistical \SI{50}{\percent} chance.
From \cref{fig:results:information}, one can see that this happens significantly earlier for omnitrees than octrees, as indicated by the numbers for the resolution $N$.
As a drastic example, if the function is $1$ only in one corner of a cube, then the omnitree can avoid generating neighboring cubes that would all be zero---it will generate one large zero-valued cuboid that spans one half of the cube, and one that spans a quarter, and one that spans one eighth of the cube, in short:
The $3$-d omnitree only needs three areas of value $0$ where the octree needs seven.
This effect will also become visible when looking closely at the tetrahedron shape in the next section.

\subsection{A Closer Look at Select Shapes in Three and Four Dimensions}\label{subsec:results:select_3d}

\begin{figure}[!htbp]
    \centering
    \begin{subfigure}[t]{0.47\textwidth}
        \centering
\pgfplotstableread[
  col sep=comma, header=true
  ]{special/l1_errors_s512_by_id.csv}\errorsTable

\begin{tikzpicture}
  \begin{axis}[
      ymode=log,
      xmode=log,
  		errorplot,
      ymin=3e-4,
      ymax=0.5,
      xmin=12,
      xmax=32800,
      minmaxbars,
      legend pos=south west,
      grid=major,
      table/x expr=\thisrow{allowed_tree_boxes},
      powtwoxlabels,
      table/search path={.,data/special/,../data/special/},
    ]
    \addplot[octree filtered, thingitet filtered, minmaxbars] table[] {l1_errors_s512_by_id.csv} coordinate (lastocttet) {};
    \addplot[omnione filtered, thingitet filtered] table[] {l1_errors_s512_by_id.csv};

    \addplot[octree filtered, thingisphere filtered] table[] {l1_errors_s512_by_id.csv} coordinate (lastoctsphere) {};
    \addplot[omnione filtered, thingisphere filtered] table[] {l1_errors_s512_by_id.csv};
    \addplot[octree filtered, thingirod filtered] table[] {l1_errors_s512_by_id.csv} coordinate (lastoctrod) {};
    \addplot[omnione filtered, thingirod filtered] table[] {l1_errors_s512_by_id.csv};
    \addplot[octree filtered, thingihilbert filtered] table[] {l1_errors_s512_by_id.csv} coordinate (lastocthilbert) {};
    \addplot[omnione filtered, thingihilbert filtered] table[] {l1_errors_s512_by_id.csv};
    \addplot[octree filtered, thingikitty filtered] table[] {l1_errors_s512_by_id.csv} coordinate (lastoctkitty) {};
    \addplot[omnione filtered, thingikitty filtered] table[] {l1_errors_s512_by_id.csv};
    \addplot[octree filtered, thingicube filtered] table[] {l1_errors_s512_by_id.csv};
    \addplot[omnione filtered, thingicube filtered] table[] {l1_errors_s512_by_id.csv};
    \coordinate (cube) at (rel axis cs:0.4,0.03);
    \legend{
      Octree,
      Omnitree
    }
\end{axis}
\node [thingitet,thingilabel] at (lastocttet){Tetrahedron};
\node[thingisphere,thingilabel] at (lastoctsphere){Sphere};
\node[thingirod,thingilabel] at (lastoctrod){Rod};
\node[thingihilbert,thingilabel] at (lastocthilbert){Hilbert Cube};
\node[thingikitty,thingilabel] at (lastoctkitty){Cat};
\node[thingicube,thingilabel] at (cube){Cube: all $= 0$ $\downarrow$ };
\end{tikzpicture}
        \caption{
            $L_1$ errors for select objects at various resolutions $N$ in three dimensions: \\Tetrahedron, Sphere, Cube, Rod, Cat, Hilbert Cube.
        }
        \label{fig:results:errors_special}
    \end{subfigure}
    \hfill
    \begin{subfigure}[t]{0.47\textwidth}
        \centering
        \begin{tikzpicture}
  \begin{axis}[
      ymode=log,
      xmode=log,
  		errorplot4d,
      ymin=2e-3,
      ymax=0.5,
      xmin=12,
      xmax=32800,
      minmaxbars,
      legend pos=south west,
      grid=major,
      table/x expr=\thisrow{allowed_tree_boxes},
      powtwoxlabels,
      table/search path={.,data/special_temporal/,../data/special_temporal/},
    ]
    \addplot[octree filtered, thingitet filtered, minmaxbars] table[] {l1_errors_s512_by_id.csv} coordinate (lastocttet) {};
    \addplot[omnione filtered, thingitet filtered] table[] {l1_errors_s512_by_id.csv} coordinate (lastomnitet) {};
    \addplot[octree filtered, thingisphere filtered] table[] {l1_errors_s512_by_id.csv} coordinate (lastoctsphere) {};
    \addplot[omnione filtered, thingisphere filtered] table[] {l1_errors_s512_by_id.csv};
    \addplot[octree filtered, thingirod filtered] table[] {l1_errors_s512_by_id.csv} coordinate (lastoctrod) {};
    \addplot[omnione filtered, thingirod filtered] table[] {l1_errors_s512_by_id.csv} coordinate (lastomnirod) {};
    \addplot[octree filtered, thingihilbert filtered] table[] {l1_errors_s512_by_id.csv} coordinate (lastocthilbert) {};
    \addplot[omnione filtered, thingihilbert filtered] table[] {l1_errors_s512_by_id.csv};
    \addplot[octree filtered, thingikitty filtered] table[] {l1_errors_s512_by_id.csv} coordinate (lastoctkitty) {};
    \addplot[omnione filtered, thingikitty filtered] table[] {l1_errors_s512_by_id.csv} coordinate (lastomnikitty) {};
    \addplot[octree filtered, thingicube filtered] table[] {l1_errors_s512_by_id.csv} coordinate (lastoctcube) {};
    \addplot[omnione filtered, thingicube filtered] table[] {l1_errors_s512_by_id.csv};
    \legend{
      Octree,
      Omnitree
    }
\end{axis}
\node [thingitet,thingilabel] at (lastomnitet){Tetrahedron};
\node[thingisphere,thingilabel] at (lastoctsphere){Sphere};
\node[thingirod,thingilabel] at (lastomnirod){Rod};
\node[thingihilbert,thingilabel] at (lastocthilbert){Hilbert Cube};
\node[thingikitty,thingilabel] at (lastomnikitty){Cat};
\node[thingicube,thingilabel] at (lastoctcube){Cube};
\end{tikzpicture}
        \caption{
            $L_1$ errors for the same objects and resolutions as in \cref{fig:results:errors_special}, but in the time-dependent four-dimensional setting.
        }
        \label{fig:results:errors_special_4d}
    \end{subfigure}
    \caption{
        $L_1$ errors evaluated for select objects at various resolutions $n$ in the $3$-d or $4$-d discretization.
        For \cref{fig:results:errors_special}, adaptation and evaluation were run in five independent instances;
        The error bars indicate the maximum and minimum values, but they are mostly hidden by the markers indicating the mean.
        This serves to illustrate that there is little variation due to the Monte-Carlo approach (only in the order of \num{e-4}), which justifies omitting further statistics on the aggregated samples in \cref{fig:results:errors10k}.
    }
\end{figure}

\begin{figure}[!htbp]
    \centering
    \newcommand{\thingiID}{0}
    \includestandalone[mode=tex,width=0.99\textwidth]{animated_gifs}
    \renewcommand{\thingiID}{1}
    \includestandalone[mode=tex,width=0.99\textwidth]{animated_gifs}
    \renewcommand{\thingiID}{187279}
    \includestandalone[mode=tex,width=0.99\textwidth]{animated_gifs}
    \caption{
        Original triangle models and their octree and omnitree refinements: Tetrahedron (related to $l_1$'s unit sphere), Sphere ($l_2$'s unit sphere), Cube ($l_\infty$'s unit sphere).
        Animations available with \href{https://tex.stackexchange.com/questions/235139/using-the-animate-package-without-adobe}{various pdf readers}:
        Build-up of discretizations from $N=16$ up to $N=\num{32768}$.
        If only one image is displayed, it is for $N=\num{32768}$.
        The corresponding $L_1$ errors are plotted in \cref{fig:results:errors_special}.
    } 
    \label{fig:results:special_3d_simple}
\end{figure}

\begin{figure}[!htbp]
    \centering
    \newcommand{\thingiID}{2}
    \includestandalone[mode=tex,width=0.99\textwidth]{animated_gifs}
    \renewcommand{\thingiID}{100349}
    \includestandalone[mode=tex,width=0.99\textwidth]{animated_gifs}
    \renewcommand{\thingiID}{53750}
    \includestandalone[mode=tex,width=0.99\textwidth]{animated_gifs}
    \caption{
        Original triangle models and their octree and omnitree refinements: Rod, Cat, Hilbert Cube.
        Animations available with \href{https://tex.stackexchange.com/questions/235139/using-the-animate-package-without-adobe}{various pdf readers}:
        Build-up of discretizations from $N=16$ up to $N=\num{32768}$.
        If only one image is displayed, it is for $N=\num{32768}$.
        The corresponding $L_1$ errors are plotted in \cref{fig:results:errors_special}.
    }
    \label{fig:results:special_3d_complex}
\end{figure}

To understand the range of outcomes in the errors, this section analyzes some shapes in more detail.
This includes shapes commonly used in modeling and simulation (tetrahedron, sphere, cube) and three more complex shapes that visually help explain the differences between octree and omnitree error measures (rod, cat, Hilbert cube).
\Cref{fig:results:errors_special} shows the errors attained for the different objects, and \cref{fig:results:special_3d_simple,fig:results:special_3d_complex} show the octree and omnitree approximations of the shapes.

The \emph{tetrahedron} (or 3-d simplex) is closely related to the unit sphere of the $l_1$ norm, and therefore of special interest for scientific computing.
Since the structure of this object is very isotropic, one would not expect a fundamental advantage of the omnitree over the octree.
However, the omnitree error exhibits a constant factor $ < 1$ when compared to the octree.
This is due to its ability to store such areas where there is only one corner of a different value, as hinted in the last section.
For instance, one can compare the (grey shaded) edges of the cuboids in the two discretizations in the first figure in \cref{fig:results:special_3d_simple}:
While the octree has to use cubes everywhere, we see cuboids of aspect ratio $2:1:1$ at the top layer of the omnitree discretization, and this pattern keeps repeating at every scale along the diagonal triangular interface of the filled and empty regions.
On the surface of this diagonal interface, \cref{fig:results:special_3d_simple} shows an emergent pattern of the Sierpinski triangle~\cite{baderSpaceFillingCurves2013,magroneSierpinskisCurveBeautiful2020} for finer resolutions.

The \emph{sphere} is itself the unit sphere of the $l_2$ norm (sometimes called RMS(E) norm).
The difference between octree and omnitree errors is larger than for the tetrahedron, because now there is anisotropy in some areas, especially in areas where the cube almost touches the bounding box.
This advantage due to this shape anisotropy is slowly increasing for higher resolutions.

As a last norm-inspired shape, we look at the \emph{cube} (\href{https://web.archive.org/web/20250623003312/https://ten-thousand-models.appspot.com/detail.html?file_id=187279}{Thingi 187279}) as the unit sphere of the maximum norm.
Both octree and omnitree can immediately perfectly resolve it with just a single cuboid.{}
The cube (and some other objects) are the reason why the minimum value for the errors in \cref{fig:results:errors10k} is equal to $0$ at all resolution levels.
Accordingly, it is one of the few examples among the $3$-d objects where the octree's total storage cost was lower to obtain the same error compared to the omnitree, cf. \cref{fig:results:memory}.

In order to analyze the error behavior for mixed-dimension curved surfaces, we further construct a cylinder with radius \num{0.05} and length \num{1}, rotated by $\frac{\pi}{4}$ around axis $(1, 1, 0)^T$, and call it \emph{rod}.
Its initial error is relatively low, due its volume being rather low (even if the approximation is $0$ everywhere, it is already relatively good, and a few levels of resolution are required to even introduce cuboids that have non-zero values in $\hat{g}$).
For finer resolutions, we again observe a slowly growing factor in the errors, but since there is relatively little anisotropy, this factor grows slower than for the sphere. 

One object that combines different kinds of smoothness, anisotropy, and scales is the \emph{cat} (\href{https://web.archive.org/web/20250509051543/https://ten-thousand-models.appspot.com/detail.html?file_id=100349}{Thingi 100349}).
In the error plot, \cref{fig:results:errors_special}, its omnitree and octree errors quickly obtain a relatively constant difference factor, with the octree lagging behind the omnitree error by approximately two horizontal ticks;
Thus the octree needs about four times the number of cuboids to achieve the same error as the omnitree.
The animation in \cref{fig:results:special_3d_complex} illustrates what this means for $3$-d shapes:
The approximate cat \enquote{grows} ears and a tail much quicker with the omnitree;
The first filled tail cuboids for the omnitree cat appear at a resolution of \num{512}, but only at \num{2048} for the octree--by which time the omnitree has already connected all tail sections.

Lastly, we investigate one of the objects that contributed the highest errors in \cref{fig:results:errors10k}, the \emph{Hilbert-cube} (\href{https://web.archive.org/web/20250509051540/https://ten-thousand-models.appspot.com/detail.html?file_id=53750}{Thingi 53750}).
As an object based on the space-filling Hilbert curve, it exhibits almost uniform density of filled volume at the coarser resolution levels.
As a result, the octree discretization can only start producing filled cubes at $N=1024$, when the cubes are starting to have a side length of $\frac{1}{16}$ and less.
For the omnitree, side lengths of $\frac{1}{16}$ already appear at $N=16$, and filled slabs along the denser areas are placed earlier, allowing for an ever widening gap between the approximation errors.


\begin{figure}[!htbp]
    \centering
    \newlength{\thirdlinewidth}
    \newlength{\twothirdlinewidth}
    \newcommand{\imgpath}{./data/special_temporal/}
    \newcommand{\thingiID}{0}
    \includestandalone[mode=tex,width=0.99\textwidth]{rotating_gifs}
    \renewcommand{\thingiID}{1}
    \includestandalone[mode=tex,width=0.99\textwidth]{rotating_gifs}
    \renewcommand{\thingiID}{187279}
    \includestandalone[mode=tex,width=0.99\textwidth]{rotating_gifs}
    \caption{
        Original time-dependent triangle models and their octree and omnitree refinements at $N=\num{32768}$: Tetrahedron, Sphere, Cube.
        Animations consist of \num{64} frames, spanning the periodic time interval $[0,1)$.
        If only one image is displayed, it is for $t=\frac{1}{128}$.
        The corresponding $L_1$ errors are plotted in \cref{fig:results:errors_special_4d}.
    } 
    \label{fig:results:special_4d_simple}
\end{figure}

\begin{figure}[!htbp]
    \centering
    \newcommand{\imgpath}{./data/special_temporal/}
    \newcommand{\thingiID}{2}
    \includestandalone[mode=tex,width=0.99\textwidth]{rotating_gifs}
    \renewcommand{\thingiID}{100349}
    \includestandalone[mode=tex,width=0.99\textwidth]{rotating_gifs}
    \renewcommand{\thingiID}{53750}
    \includestandalone[mode=tex,width=0.99\textwidth]{rotating_gifs}
    \caption{
        Original time-dependent triangle models and their octree and omnitree refinements at $N=\num{32768}$: Rod, Cat, Hilbert Cube.
        Animations consist of \num{64} frames, spanning the periodic time interval $[0,1)$.
        If only one image is displayed, it is for $t=\frac{1}{128}$.
        The corresponding $L_1$ errors are plotted in \cref{fig:results:errors_special_4d}.
    }
    \label{fig:results:special_4d_complex}
\end{figure}

\Cref{fig:results:errors_special_4d} shows the errors for a similar analysis of the same objects;
The difference is that they were made four-dimensional by a time-dependent rotation along the axis $(1,1,1)^T$.
One full rotation is performed from time $t=0$ to $t=1$, and the objects are re-scaled to fit into the unit hypercube at every time step.
The results for resolution level $N=\num{32768}$ are shown in \cref{fig:results:special_4d_simple,fig:results:special_4d_complex}.

For the \emph{tetrahedron} and \emph{cube}, we now observe that the resolution in both time and space is much lower for the octree than the omnitree, because regions of fine resolution need to consider all four dimensions at once.
This is visible by the high number of empty, seemingly \enquote{unused} cubes around the octree-discretized object, which do not appear for the omnitree.

The same effect is even more drastic for the \emph{sphere}, where the octree discretization needs to spend much resources on temporal resolution.
The octree discretization has to resolve the time dimension just as finely as the spatial dimensions along the sphere surface, even though this results in barely any change over time.
The omnitree, by comparison, can exploit the fact that the sphere interior is constant over time, and spends all rectangles on the spatial resolution, leading to a drastic divergence of the errors in \cref{fig:results:errors_special_4d}---approximately a factor of \num{10} for $N =\num{32768}$.{}

For the \emph{rod} and \emph{cat}, the time-space tradeoff available to the omnitree works generally in the same way as for the tetrahedron and cube.
However, due to the fact that there are larger stable regions (the object is present in the middle of the domain throughout the time, most of the corners are always empty), and the relative isotropy of the regions in between, the difference in the errors is not as drastic as for the previous objects.

Lastly, the \emph{Hilbert cube} provides again a very interesting observations:
Considering the errors in \cref{fig:results:errors_special_4d}, a resolution of $N=\num{32768}$ is not nearly sufficient to reduce the error at all.
The octree in the lowest-placed animation in \cref{fig:results:special_4d_complex} fills only a few hypercubes with $1$ values, while the omnitree can nearly represent the Hilbert cube when it is (nearly) axis-aligned; 
This is the case at times $t=0 (=1), t=\frac{1}{3}$, and $t=\frac{2}{3}$.
However, due to the time dependent rotation with rescaling, the time when this approximation is close to the true solution is very short, and many more rectangles would be needed to reduce the error significantly (which was out of the scope of this work).
Conclusively, even if the omnitree can represent objects more faithfully than the octree, only switching to omnitree discretization may not be sufficient to obtain a good approximation of the true function.
Compared to octrees, it can be even more important to transform the problem in a way that aligns it more with the coordinate axes.

Overall, these specially selected objects are relatively simple, and most of them (tetrahedron, sphere, rod) exhibit their important features already at the lowest resolution levels.
This is not the case for the majority of objects in \thingitenk{}, as they are designed to be printed and composed with other parts (i.e., they will often be flat or long, axis-aligned by construction, and have small features for grooves and joints).
For example, one of the objects that see the most benefit from omnitrees (\href{https://web.archive.org/web/20250731030202/https://ten-thousand-models.appspot.com/detail.html?file_id=55177}{Thingi 551777}) is part of a two-material (\enquote{dual-extrusion}) cupcake model and consists of a flat disk at the bottom and a feature-rich cake part at the top, for interlaced printing with the other color. 
For these reasons, the median error difference observed in \cref{fig:results:errors10k} is higher than for the better-explainable shapes discussed here.{}

\subsection{F25 Showcase}\label{subsec:aeroplane}
\begin{figure}[!ht]
    \centering
\pgfplotstableread[
  col sep=comma, header=true
  ]{f25/l1_errors_s512.csv}\errorsTable
\def\axisdefaultwidth{10cm}
\def\axisdefaultheight{6cm}
\begin{tikzpicture}
  \begin{axis}[
      ymode=log,
      xmode=log,
  		errorplot,
      ymin=3e-4,
      ymax=0.01, 
      xmin=12,
      xmax=262200,
      legend pos=south west,
      grid=major,
      table/x expr=\thisrow{allowed_tree_boxes},
      table/y expr=\thisrow{l1error},
      powtwoxlabels,
      log ticks with fixed point,
      table/search path={.,data/f25/,../data/f25/},
    ]
    \addplot[octree filtered, thingiplane] table[] {l1_errors_s512.csv} coordinate (lastoctplane) {};
    \addplot[omnione filtered, thingiplane] table[] {l1_errors_s512.csv};
    \legend{
      Octree,
      Omnitree
    }
\end{axis}
\node[thingiplane,thingilabel] at (lastoctplane){F25 Plane Model};
\end{tikzpicture}\\
    \includegraphics[trim=10pt 400pt 0pt 400pt, clip,width=0.98\textwidth]{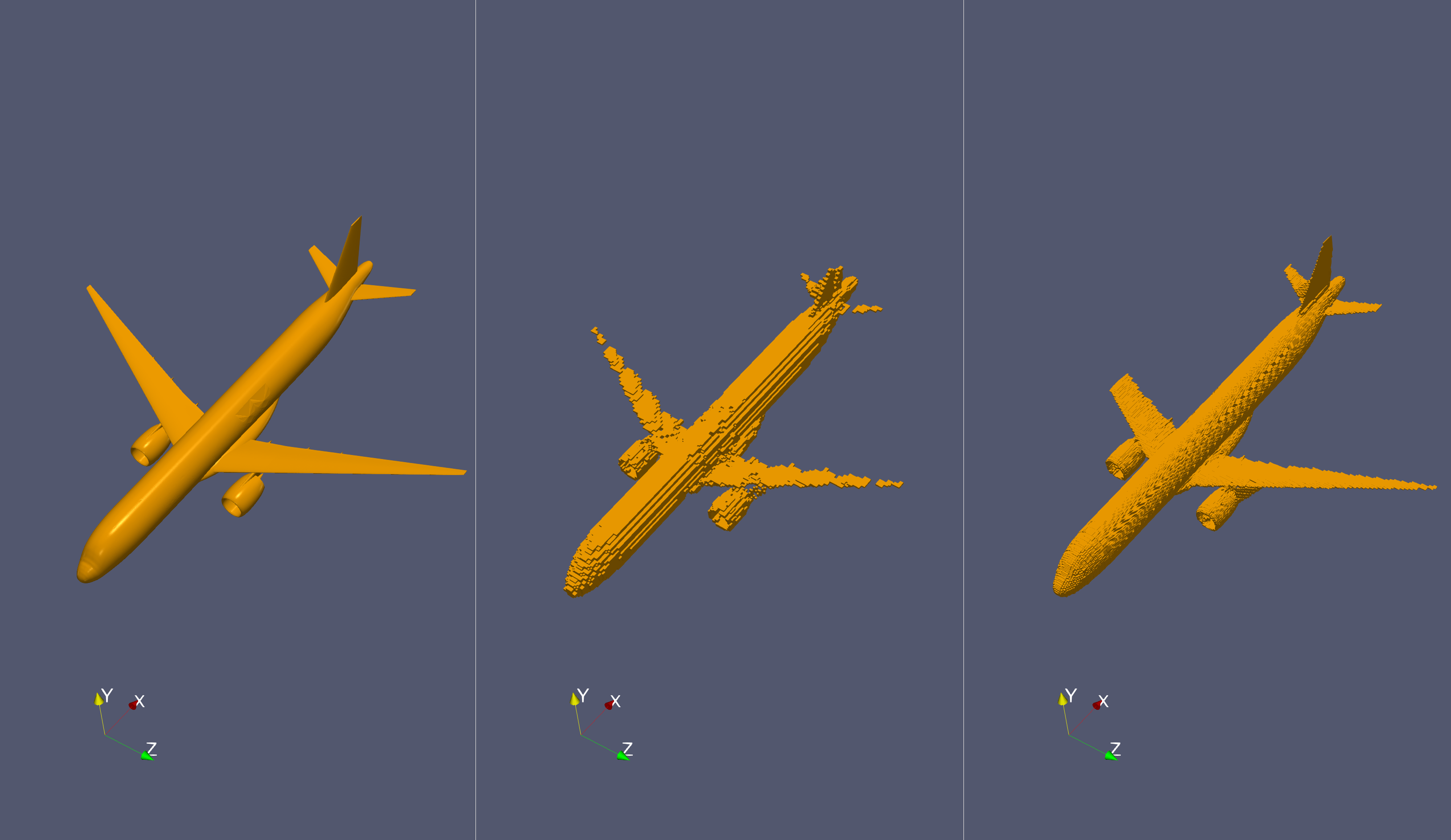}
    \caption{
        Comparison of octree and omnitree Monte Carlo $L_1$ errors for $N=\numrange{16}{262144}$ on the F25 short-to-medium range research baseline aircraft~\cite{wohlerEstablishingDLRF25Research2024,DLRF25} (upper image).
        Original triangle model of the F25 (lower left), and models of its octree (lower mid) and omnitree refinements (lower right) at $N=\num{32768}$ (an intermediate resolution allowing for a good visual comparison).
        At $N=\num{32768}$, the difference between octree and omnitree errors amounts to a factor of \num[round-mode=places,round-precision=1]{2.398373984}$\times$.
    }
    \label{fig:results:f25_3d}
\end{figure}

Lastly, we use the F25 short-to-medium range research baseline aircraft~\cite{wohlerEstablishingDLRF25Research2024,DLRF25} as a showcase for practical applications at very fine resolutions.
The F25 model (after slight editing for simplification) contains \num{273481} vertices and \num{546902} triangles.
The refinement algorithm as described in \cref{sec:eval} was executed on the F25 triangle model for resolutions from $N=\num{16}$ to \num{262144} , and the resulting octree and omnitree discretization errors are compared in \cref{fig:results:f25_3d}.
The errors obtained for the octree's resolution levels \numrange{1024}{16384} roughly correspond to the errors at the omnitree's resolution levels \numrange{64}{1024}, indicating a \num{16}$\times$ advantage for the omnitree in that regime.
We also provide a visual comparison of the approximate aircraft models to the original model in \cref{fig:results:f25_3d}.
Due to the probabilistic nature of the refinement algorithm, the omnitree refinement happens to take no sample inside the outer part of the right wing early on, such that this part of the aircraft is not resolved at all in the omnitree discretization.
This illustrates that practical simulation applications should rather use dimension-adaptive versions of existing intersection-based algorithms~\cite{losassoSimulatingWaterSmoke2004,mirjaliliInterfacecapturingMethodsTwophase2017} (and not the information-theory inspired sensitivity indices used for this study).
Accordingly, the errors get closer for the finer resolutions, since the relative influence of the partly missing wing increases gradually.
Another limitation is that the number of error samples $n_e = \num{262144}$ is becoming too small at finer resolutions, and the errors become more noisy and less reliable as a result.{}
Given this caveat, one can see both in the error plot and the visualizations that the omnitree is able to resolve the F25 aircraft much better at the same resolution than the octree.
In practice, such a model could be used as mask for spatial occupancy in an omnitree-based rectilinear adaptive CFD method, for example in conjunction with level-sets or other sharp-interface resolution methods~\cite{mirjaliliInterfacecapturingMethodsTwophase2017}.

\section{Implications for Applications}\label{sec:discussion}
The omnitree approach to dyadic adaptivity derives its benefits from the fact that current computers store their results in binary form.
One instance of this is the unique compact bitstring representation of the tree.
This representation is already quite memory efficient and has the potential for further compression, as most operations on this bitstring operate in a streaming fashion.
In our Python implementation of omnitree refinement~\cite{bleifreiFreifrauvonbleifreiDyAda2025} and the shape representation code used for this work~\cite{bleifreiFreifrauvonbleifreiThingies_with_omnitrees2025}, the bitstring descriptor is stored separately from the function data $\hat{g}$.

Even though this work does not consider computational performance, we would like to mention another important binary aspect:
The natural nesting of the space-filling Z curve with the binary numbers (as explored by e.g. Connor and Kumar~\cite{connorFastConstructionKnearest2010}), can allow for high cache efficiency for localized operations such as stencils/convolutions and is directly applicable to omnitrees.
At the same time, indexing into the omnitree in its current implementation needs linear time in $N$ and requires many bit and integer operations, which can be slow on architectures geared mostly towards floating point operations.
This can be addressed by indexing the coarsest levels of the omnitree with hashmap- or pointer-based tree data structures.
The Z curve nesting will further allow for omnitrees to be parallelized in the well-established way practiced, for instance, in the \texttt{p4est}~\cite{bursteddeP4estScalableAlgorithms2011} and \texttt{t8code}~\cite{holkeT8codeModularAdaptive2025} high-performance AMR frameworks.
To achieve this, further research into two-to-one balancing and efficient iteration of omnitrees will be necessary.



As derived in \cref{subsec:storage_convergence}, omnitrees will be particularly advantageous in scenarios that exhibit high dimensionalities and high anisotropy:
While the storage cost for the tree itself will increase linearly with the dimensionality $d$, the reduction of the error for anisotropic problems can be exponential in $d$ through the higher order of convergence.
Compared to octrees, the relative increase in storage costs is particularly small if the data vector $\hat{g}$ uses long data types such as single- and double-precision floating points (32 and 64 bits, respectively)---if the tree data is vanishingly small compared to the function data, then enlarging it by a moderate factor $d$ will still result in a relatively cheap tree data structure.
The reduction of the error could in fact be validated in \cref{sec:results}, where the convergence rate was increased by a factor of \num[round-mode=places,round-precision=2]{1.348535914} for the median errors and \num[round-mode=places,round-precision=2]{1.526316811} for the mean errors on a moderately anisotropic $3$-d problem.
The results of \cref{subsec:results:select_3d} provide a first indication that the omnitree's advantage will in fact be stronger for higher dimensionalities.
We could also observe that the lower errors immediately outweigh the increased storage costs compared to octrees (\cref{fig:results:memory}).
Notably, these results were established for the problem of $3$-d binary shape representation: moderate dimensionality, moderate anisotropy, and the smallest possible data type used for the data vector.
Accordingly, we expect omnitree applications to be useful in all areas where currently octrees or bintrees are used---including but not limited to compression, vector databases, computer graphics, simulations, and machine learning.{}
And we expect the benefits to be the most evident for problems that exhibit higher dimensionalities, high anisotropy, and long data types;
Accordingly, omnitrees may make AMR accessible to fields that could not utilize it so far, such as plasma microturbulence simulations.

\section{Conclusion}\label{sec:conclusion} 

This paper serves as an introduction of omnitrees as widely applicable anisotropic adaptive data structures.
It discusses how omnitrees generalize octrees as well as bintrees, and how they relate to AMM.
While certain aspects of omnitrees have been demonstrated before, this work (and the accompanying Python implementation~\cite{bleifreiFreifrauvonbleifreiDyAda2025}) is the first to make them practically usable for arbitrary dimensionalities.
This is achieved through a novel tree formulation enabling their storage as a compact linearized data structure.
Through theoretical analysis, we have shown that that omnitrees can increase the order of convergence by up to a factor of $d$, the dimensionality, compared to octrees.
This effect compounds over all orders of magnitude and thus offsets the moderate increase of storage costs for the tree descriptor.
Based on more than four thousand $3$-d objects, these claims were validated on the problem of shape representation.
We observe that the omnitree representation using \num{8192} cuboids per object achieved a \num[round-mode=places,round-precision=1]{10.135638298}${\times}$ lower median error than the octree representation at the same equivalent resolution.
Using a binary function as solution provided insights into how the information density in the discretized function is connected to error convergence.
Through close examination of a few select objects, we were able to visually explain the underlying effects and further extend them to a time-dependent $4$-d problem on rotated shapes.
We argue that every problem currently addressed with octrees or bintrees could potentially benefit from omnitrees, and for some problems, AMR may become tractable through omnitrees for the first time.
As concrete next steps, we aim to investigate how coarsening can be formulated and efficiently implemented in the context of omnitrees.
A driver to this pursuit are applications which expect a minimal resolution in which the data should be represented exactly, which omnitrees could achieve by coarsening from this minimal resolution.
This, in turn, will enable further comparison to state-of-the-art formats for structured data such as OpenVDB~\cite{musethOpenVDBOpensourceData2013}.

\section*{Acknowledgements}
We would like to thank Malte Brunn for helpful discussions in an early phase of the project, and Alexandre Bardakoff for relentless praise and support.
We would like to thank Neda Ebrahimi Pour for suggesting the use of the F25 plane model.
The development of the DLR-F25 configuration is funded by the German Federal Ministry for Economic Affairs and Climate Action (BMWK) as part of the LuFo VI-2 project VirEnfREI (funding reference: 20X2106B).

\section*{Declaration of Generative AI and AI-assisted technologies in the writing process}

Statement: During the preparation of this work the authors used Microsoft Copilot in order to edit the manuscript text.
After using this tool/service, the authors reviewed and edited the content as needed and take full responsibility for the content of the publication.

\section*{Reproducibility References}
Code used for this work includes omnitree refinement~\cite{bleifreiFreifrauvonbleifreiDyAda2025} and shape approximation based on sensitivity analysis~\cite{bleifreiFreifrauvonbleifreiThingies_with_omnitrees2025}; both are available through github under GPL-3.0.
Intermediate data produced are available through a research data repository~\cite{pollingerReproducibilityDataBeauty2025}.

\ifthenelse{\boolean{arxiv}}{
    \printbibliography

@book{baderSpaceFillingCurves2013,
  title = {Space-{{Filling Curves}}},
  author = {Bader, Michael},
  year = {2013},
  series = {Texts in {{Computational Science}} and {{Engineering}}},
  volume = {9},
  publisher = {Springer},
  location = {Berlin, Heidelberg},
  doi = {10.1007/978-3-642-31046-1},
  isbn = {978-3-642-31045-4}
}

@article{bhatiaAMMAdaptiveMultilinear2022,
  title = {{{AMM}}: {{Adaptive Multilinear Meshes}}},
  shorttitle = {{{AMM}}},
  author = {Bhatia, Harsh and Hoang, Duong and Morrical, Nate and Pascucci, Valerio and Bremer, Peer-Timo and Lindstrom, Peter},
  date = {2022-06},
  year = {2022},
  journal = {IEEE Transactions on Visualization and Computer Graphics},
  volume = {28},
  number = {6},
  pages = {2350--2363},
  issn = {1941-0506},
  doi = {10.1109/TVCG.2022.3165392},
  eventtitle = {{{IEEE Transactions}} on {{Visualization}} and {{Computer Graphics}}}
}

@incollection{domelSplitflowProgress3D2000,
  title = {Splitflow - {{Progress}} in {{3D CFD}} with {{Cartesian}} Omni-Tree Grids for Complex Geometries},
  booktitle = {38th {{Aerospace Sciences Meeting}} and {{Exhibit}}},
  author = {Domel, Neal and Karman, Steve},
  year = {2000},
  publisher = {{American Institute of Aeronautics and Astronautics}},
  doi = {10.2514/6.2000-1006},
}

@misc{haverkortSixteenSpacefillingCurves2018a,
  title = {Sixteen Space-Filling Curves and Traversals for d-Dimensional Cubes and Simplices},
  author = {Haverkort, Herman},
  date = {2018-09-12},
  year = {2018},
  eprint = {1711.04473},
  eprinttype = {arXiv},
  eprintclass = {cs},
  pubstate = {prepublished}
}

@inproceedings{ogawaAdaptiveCartesianMesh2003,
  title = {An {{Adaptive Cartesian Mesh Flow Solver Based}} on the {{Tree-data}} with {{Anisotropic Mesh Refinement}}},
  booktitle = {Computational {{Fluid Dynamics}} 2002},
  author = {Ogawa, Takanobu},
  editor = {Armfield, Steve W. and Morgan, Patrick and Srinivas, Karkenahalli},
  year = {2003},
  pages = {453--458},
  publisher = {Springer},
  location = {Berlin, Heidelberg},
  doi = {10.1007/978-3-642-59334-5_67},
  isbn = {978-3-642-59334-5},
  langid = {english}
}

@inproceedings{ogawaNumericalAlgorithmBased2009,
  title = {A {{Numerical Algorithm Based}} on a {{Tree Data}} for an {{Anisotropically Adaptive Cartesian Mesh}}},
  author = {Ogawa, Takanobu},
  year = {2009},
  pages = {2015--2021},
  publisher = {American Society of Mechanical Engineers Digital Collection},
  doi = {10.1115/FEDSM2003-45532},
  eventtitle = {{{ASME}}/{{JSME}} 2003 4th {{Joint Fluids Summer Engineering Conference}}},
  langid = {english}
}

@incollection{ogawaParallelizationAdaptiveCartesian2003,
  title = {Parallelization of an {{Adaptive Cartesian Mesh Flow Solver Based}} on the $2^{{N}}$-Tree {{Data Structure}}},
  booktitle = {Parallel {{Computational Fluid Dynamics}} 2002},
  author = {Ogawa, Takanobu},
  editor = {Matsuno, K. and Ecer, A. and Satofuka, N. and Periaux, J. and Fox, P.},
  year = {2003},
  pages = {441--448},
  publisher = {North-Holland},
  location = {Amsterdam},
  doi = {10.1016/B978-044450680-1/50056-5},
  isbn = {978-0-444-50680-1}
}

@article{sametQuadtreeRelatedHierarchical1984,
  title = {The {{Quadtree}} and {{Related Hierarchical Data Structures}}},
  author = {Samet, Hanan},
  year = {1984},
  journal = {ACM Computing Surveys (CSUR)},
  volume = {16},
  number = {2},
  pages = {187--260},
  doi = {10.1145/356924.356930}
}

@incollection{sangApplicationOmnitreeAdaptive2003,
  title = {Application of an {{Omni-tree Adaptive}} and {{Hybrid Cartesian Grid Method}}},
  booktitle = {41st {{Aerospace Sciences Meeting}} and {{Exhibit}}},
  author = {Sang, Wei and Li, Feng and E, Q.},
  year = {2003},
  publisher = {{American Institute of Aeronautics and Astronautics}},
  doi = {10.2514/6.2003-1239},
}

@article{sangComparisonOctreeOmniTree2013,
  title = {Comparison of {{Octree}} and {{Omni-Tree Cartesian Grid}} for {{Civil-Plane High-Lift Model Simulations}}},
  author = {Sang, Weimin and Shi, Yu},
  year = {2013},
  journal = {Journal of Aircraft},
  volume = {50},
  number = {4},
  pages = {1099--1105},
  publisher = {{American Institute of Aeronautics and Astronautics}},
  issn = {0021-8669},
  doi = {10.2514/1.C032044}
}

@article{sangNumericallyAnalyzingMore2011,
  title = {Numerically Analyzing More Efficiently High-Lift Aerodynamics of Wing/Body Model with Omni-Tree {{Cartesian}} Grids},
  author = {Sang, Weimin and Yu, Jiangang},
  year = {2011},
  journal = {Aerospace Science and Technology},
  volume = {15},
  number = {5},
  pages = {375--380},
  issn = {1270-9638},
  doi = {10.1016/j.ast.2010.09.001}
}

@incollection{sangOmnitreeAdaptiveCartesian2003,
  title = {Omni-Tree and {{Adaptive Cartesian Hybrid Grid Method}} in {{Steady}} and {{Unsteady Flows}}},
  booktitle = {21st {{AIAA Applied Aerodynamics Conference}}},
  author = {Sang, Wei and Li, Feng},
  year = {2003},
  publisher = {{American Institute of Aeronautics and Astronautics}},
  doi = {10.2514/6.2003-4078},
}

@misc{zhouThingi10KDataset100002016,
  title = {{{Thingi10K}}: {{A Dataset}} of 10,000 {{3D-Printing Models}}},
  shorttitle = {{{Thingi10K}}},
  author = {Zhou, Qingnan and Jacobson, Alec},
  date = {2016-07-02},
  year = {2016},
  eprint = {1605.04797},
  eprinttype = {arXiv},
  eprintclass = {cs},
  doi = {10.48550/arXiv.1605.04797},
  pubstate = {prepublished}
}

@inproceedings{zhangAdaptivePatchingHighresolution2024a,
  title = {Adaptive {{Patching}} for {{High-resolution Image Segmentation}} with {{Transformers}}},
  booktitle = {{{SC24}}: {{International Conference}} for {{High Performance Computing}}, {{Networking}}, {{Storage}} and {{Analysis}}},
  author = {Zhang, Enzhi and Lyngaas, Isaac and Chen, Peng and Wang, Xiao and Igarashi, Jun and Huo, Yuankai and Munetomo, Masaharu and Wahib, Mohamed},
  year = {2024},
  pages = {1--16},
  publisher = {IEEE},
  url = {https://ieeexplore.ieee.org/abstract/document/10793226/},
  urldate = {2025-04-22}
}

@article{saltelliMakingBestUse2002,
  title = {Making Best Use of Model Evaluations to Compute Sensitivity Indices},
  author = {Saltelli, Andrea},
  date = {2002-05-15},
  year = {2002},
  journal = {Computer Physics Communications},
  volume = {145},
  number = {2},
  pages = {280--297},
  issn = {0010-4655},
  doi = {10.1016/S0010-4655(02)00280-1},
  langid = {english}
}

@book{saltelliGlobalSensitivityAnalysis2008,
  title = {Global Sensitivity Analysis: The Primer},
  shorttitle = {Global Sensitivity Analysis},
  author = {Saltelli, Andrea and Ratto, Marco and Andres, Terry and Campolongo, Francesca and Cariboni, Jessica and Gatelli, Debora and Saisana, Michaela and Tarantola, Stefano},
  year = {2008},
  publisher = {John Wiley \& Sons}
}

@article{borgonovoNewUncertaintyImportance2007,
  title = {A New Uncertainty Importance Measure},
  author = {Borgonovo, E.},
  date = {2007-06-01},
  year = {2007},
  journal = {Reliability Engineering \& System Safety},
  volume = {92},
  number = {6},
  pages = {771--784},
  issn = {0951-8320},
  doi = {10.1016/j.ress.2006.04.015}
}

@article{kawaguchiMethodBinaryPictureRepresentation1980,
  title = {On a {{Method}} of {{Binary-Picture Representation}} and {{Its Application}} to {{Data Compression}}},
  author = {Kawaguchi, Eiji and Endo, Tsutomu},
  date = {1980-01},
  year = {1980},
  journal = {IEEE Transactions on Pattern Analysis and Machine Intelligence},
  volume = {PAMI-2},
  number = {1},
  pages = {27--35},
  issn = {1939-3539},
  doi = {10.1109/TPAMI.1980.4766967}
}

@article{knowltonProgressiveTransmissionGreyscale1980,
  title = {Progressive Transmission of Grey-Scale and Binary Pictures by Simple, Efficient, and Lossless Encoding Schemes},
  author = {Knowlton, K.},
  date = {1980-07},
  year = {1980},
  journal = {Proceedings of the IEEE},
  volume = {68},
  number = {7},
  pages = {885--896},
  issn = {1558-2256},
  doi = {10.1109/PROC.1980.11754}
}

@incollection{domelTimeTensorRapidConvergence2010,
  title = {Time-{{Tensor}} for {{Rapid Convergence}} of {{CFD Solutions}}},
  booktitle = {48th {{AIAA Aerospace Sciences Meeting Including}} the {{New Horizons Forum}} and {{Aerospace Exposition}}},
  author = {Domel, Neal},
  year = {2010},
  publisher = {{American Institute of Aeronautics and Astronautics}},
  doi = {10.2514/6.2010-120},
}

@incollection{domelPropulsionAerodynamicsWorkshop2019,
  title = {Propulsion {{Aerodynamics Workshop IV Results}} for the {{Special Topic}}},
  booktitle = {{{AIAA Propulsion}} and {{Energy}} 2019 {{Forum}}},
  author = {Domel, Neal D.},
  year = {2019},
  publisher = {{American Institute of Aeronautics and Astronautics}},
  doi = {10.2514/6.2019-4020},
}

@incollection{baruzziniFundamentalChallengesMicroVanes2009,
  title = {Fundamental {{Challenges}} of {{Micro-Vanes}} and {{Micro-Ramps}} for {{High-Speed Inlet Applications}}: {{A Computational Fluid Dynamics Investigation}}},
  shorttitle = {Fundamental {{Challenges}} of {{Micro-Vanes}} and {{Micro-Ramps}} for {{High-Speed Inlet Applications}}},
  booktitle = {45th {{AIAA}}/{{ASME}}/{{SAE}}/{{ASEE Joint Propulsion Conference}} \& {{Exhibit}}},
  author = {Baruzzini, Dan and Domel, Neal and Miller, Daniel},
  year = {2009},
  publisher = {{American Institute of Aeronautics and Astronautics}},
  doi = {10.2514/6.2009-5074}
}

@misc{trimesh,
	author = {{Dawson-Haggerty et al.}},
	title = {trimesh},
	url = {https://trimesh.org/},
	version = {3.2.0},
	year = {2019},
}

@article{bentleyMultidimensionalBinarySearch1975,
  title = {Multidimensional Binary Search Trees Used for Associative Searching},
  author = {Bentley, Jon Louis},
  year = {1975},
  journal = {Commun. ACM},
  volume = {18},
  number = {9},
  pages = {509--517},
  issn = {0001-0782},
  doi = {10.1145/361002.361007}
}

@article{IwanagaSALib2022,
  title = {Toward {SALib} 2.0: {Advancing} the accessibility and interpretability of global sensitivity analyses},
  volume = {4},
  doi = {10.18174/sesmo.18155},
  journal = {Socio-Environmental Systems Modelling},
  author = {Iwanaga, Takuya and Usher, William and Herman, Jonathan},
  month = may,
  year = {2022},
  pages = {18155},
}

@article{HermanSALib2017,
  doi = {10.21105/joss.00097},
  year  = {2017},
  publisher = {The Open Journal},
  volume = {2},
  number = {9},
  author = {Jon Herman and Will Usher},
  title = {{SALib}: An open-source Python library for Sensitivity Analysis},
  journaltitle = {The Journal of Open Source Software},
  journal = {JOSS},
}

@article{ambrosianoFastMultipoleMethod1988,
  title = {The Fast Multipole Method for Gridless Particle Simulation},
  author = {Ambrosiano, John and Greengard, Leslie and Rokhlin, Vladimir},
  date = {1988-01-01},
  year = {1988},
  journal = {Computer Physics Communications},
  journaltitle = {Computer Physics Communications},
  volume = {48},
  number = {1},
  pages = {117--125},
  issn = {0010-4655},
  doi = {10.1016/0010-4655(88)90029-X}
}

@article{greengardFastAlgorithmParticle1987,
  title = {A Fast Algorithm for Particle Simulations},
  author = {Greengard, L and Rokhlin, V},
  date = {1987-12-01},
  year = {1987},
  journaltitle = {Journal of Computational Physics},
  journal = {J. Comput. Phys.},
  volume = {73},
  number = {2},
  pages = {325--348},
  issn = {0021-9991},
  doi = {10.1016/0021-9991(87)90140-9},
  url = {https://www.sciencedirect.com/science/article/pii/0021999187901409},
  urldate = {2025-06-20}
}

@misc{dengEfficientAutoregressiveShape2025a,
  title = {Efficient {{Autoregressive Shape Generation}} via {{Octree-Based Adaptive Tokenization}}},
  author = {Deng, Kangle and Liu, Hsueh-Ti Derek and Zhu, Yiheng and Sun, Xiaoxia and Shang, Chong and Bhat, Kiran and Ramanan, Deva and Zhu, Jun-Yan and Agrawala, Maneesh and Zhou, Tinghui},
  date = {2025-04-03},
  year = {2025},
  eprint = {2504.02817},
  eprinttype = {arXiv},
  eprintclass = {cs},
  doi = {10.48550/arXiv.2504.02817},
  pubstate = {prepublished},
  version = {1}
}

@article{whangOctreeRAdaptiveOctree1995,
  title = {Octree-{{R}}: An Adaptive Octree for Efficient Ray Tracing},
  shorttitle = {Octree-{{R}}},
  author = {Whang, Kyu-Young and Song, Ju-Won and Chang, Ji-Woong and Kim, Ji-Yun and Cho, Wan-Sup and Park, Chong-Mok and Song, Il-Yeol},
  date = {1995-12},
  year = {1995},
  journal = {IEEE Transactions on Visualization and Computer Graphics},
  volume = {1},
  number = {4},
  pages = {343--349},
  issn = {1941-0506},
  doi = {10.1109/2945.485621}
}

@article{hornungOctoMapEfficientProbabilistic2013,
  title = {{{OctoMap}}: An Efficient Probabilistic {{3D}} Mapping Framework Based on Octrees},
  shorttitle = {{{OctoMap}}},
  author = {Hornung, Armin and Wurm, Kai M. and Bennewitz, Maren and Stachniss, Cyrill and Burgard, Wolfram},
  date = {2013-04-01},
  year = {2013},
  journal = {Autonomous Robots},
  volume = {34},
  number = {3},
  pages = {189--206},
  issn = {1573-7527},
  doi = {10.1007/s10514-012-9321-0},
  langid = {english}
}

@article{schonOctreebasedIndexing3D2013,
  title = {Octree-Based Indexing for {{3D}} Pointclouds within an {{Oracle Spatial DBMS}}},
  author = {Schön, Bianca and Mosa, Abu Saleh Mohammad and Laefer, Debra F. and Bertolotto, Michela},
  date = {2013-02-01},
  year = {2013},
  journal = {Computers \& Geosciences},
  volume = {51},
  pages = {430--438},
  issn = {0098-3004},
  doi = {10.1016/j.cageo.2012.08.021}
}

@article{tuEtreeDatabaseorientedMethod2004,
  title = {Etree: A Database-Oriented Method for Generating Large Octree Meshes},
  shorttitle = {Etree},
  author = {Tu, T. and O’Hallaron, D. R. and López, J. C.},
  date = {2004-07-01},
  year = {2004},
  journal = {Engineering with Computers},
  volume = {20},
  number = {2},
  pages = {117--128},
  issn = {1435-5663},
  doi = {10.1007/s00366-004-0283-5},
  langid = {english}
}

@inproceedings{chenParallelNNParallelOctreebased2023,
  title = {{{ParallelNN}}: {{A Parallel Octree-based Nearest Neighbor Search Accelerator}} for {{3D Point Clouds}}},
  shorttitle = {{{ParallelNN}}},
  booktitle = {2023 {{IEEE International Symposium}} on {{High-Performance Computer Architecture}} ({{HPCA}})},
  author = {Chen, Faquan and Ying, Rendong and Xue, Jianwei and Wen, Fei and Liu, Peilin},
  date = {2023-02},
  year = {2023},
  pages = {403--414},
  issn = {2378-203X},
  doi = {10.1109/HPCA56546.2023.10070940},
  eventtitle = {2023 {{IEEE International Symposium}} on {{High-Performance Computer Architecture}} ({{HPCA}})}
}

@article{judeOctreebasedCartesianNavier2022,
  title = {An Octree-Based, Cartesian Navier–Stokes Solver for Modern Cluster Architectures},
  author = {Jude, Dylan and Sitaraman, Jayanarayanan and Wissink, Andrew},
  date = {2022-06-01},
  year = {2022},
  journaltitle = {The Journal of Supercomputing},
  journal = {J. Supercomput.},
  volume = {78},
  number = {9},
  pages = {11409--11440},
  issn = {1573-0484},
  doi = {10.1007/s11227-022-04324-7},
  langid = {english}
}

@article{bursteddeP4estScalableAlgorithms2011,
  title = {P4est: {{Scalable Algorithms}} for {{Parallel Adaptive Mesh Refinement}} on {{Forests}} of {{Octrees}}},
  shorttitle = {P4est},
  author = {Burstedde, Carsten and Wilcox, Lucas C. and Ghattas, Omar},
  date = {2011-01},
  year = {2011},
  journaltitle = {SIAM Journal on Scientific Computing},
  journal = {SIAM J. Sci. Comput.},
  volume = {33},
  number = {3},
  pages = {1103--1133},
  publisher = {{Society for Industrial and Applied Mathematics}},
  issn = {1064-8275},
  doi = {10.1137/100791634}
}

@article{holkeT8codeModularAdaptive2025,
  title = {T8code - Modular Adaptive Mesh Refinement in the Exascale Era},
  author = {Holke, Johannes and Markert, Johannes and Knapp, David and Dreyer, Lukas and Elsweijer, Sandro and Böing, Niklas and Lilikakis, Ioannis and Fussbroich, Jakob and Leistikow, Tabea and Becker, Florian and Uenlue, Veli and Albers, Ole and Burstedde, Carsten and Basermann, Achim and Hergl, Chiara and Julia, Weber and Schoenlein, Kathrin and Ackerschott, Jonas and Evgenii, Andreev and Csati, Zoltan and Dutka, Alexandra and Geihe, Benedict and Kestener, Pierre and Kirby, Andrew and Ranocha, Hendrik and Schlottke-Lakemper and Michael},
  date = {2025-02-06},
  year = {2025},
  journaltitle = {Journal of Open Source Software},
  journal = {JOSS},
  volume = {10},
  number = {106},
  pages = {6887},
  issn = {2475-9066},
  doi = {10.21105/joss.06887},
  langid = {english}
}

@inproceedings{daissPizDaintStars2019a,
  title = {From {{Piz Daint}} to the Stars: Simulation of Stellar Mergers Using High-Level Abstractions},
  shorttitle = {From Piz Daint to the Stars},
  booktitle = {Proceedings of the {{International Conference}} for {{High Performance Computing}}, {{Networking}}, {{Storage}} and {{Analysis}}},
  author = {Daiß, Gregor and Amini, Parsa and Biddiscombe, John and Diehl, Patrick and Frank, Juhan and Huck, Kevin and Kaiser, Hartmut and Marcello, Dominic and Pfander, David and Pfüger, Dirk},
  date = {2019-11-17},
  year = {2019},
  series = {{{SC}} '19},
  pages = {1--37},
  publisher = {Association for Computing Machinery},
  location = {New York, NY, USA},
  doi = {10.1145/3295500.3356221},
  isbn = {978-1-4503-6229-0}
}

@article{magroneSierpinskisCurveBeautiful2020,
  title = {Sierpinski’s Curve: {{A}} (Beautiful) Paradigm of Recursion},
  shorttitle = {Sierpinski’s Curve},
  author = {Magrone, Paola},
  year = {2020},
  journal = {Slov. Časopis Pre Geom. Graf},
  volume = {17},
  pages = {17--28},
  url = {http://ssgg.sk/G/Abstrakty/G_cisla/G33.pdf#page=17},
  urldate = {2025-06-20}
}

@article{minyardOctreePartitioningHybrid1998,
  title = {Octree Partitioning of Hybrid Grids for Parallel Adaptive Viscous Flow Simulations},
  author = {Minyard, T. and Kallinderis, Y.},
  year = {1998},
  journaltitle = {International Journal for Numerical Methods in Fluids},
  journal = {Internat. J. Numer. Methods Fluids},
  volume = {26},
  number = {1},
  pages = {57--78},
  issn = {1097-0363},
  doi = {10.1002/(SICI)1097-0363(19980115)26:1<57::AID-FLD625>3.0.CO;2-N},
  langid = {english}
}

@misc{nakasatoAstrophysicalParticleSimulations2012,
  title = {Astrophysical {{Particle Simulations}} on {{Heterogeneous CPU-GPU Systems}}},
  author = {Nakasato, Naohito and Ogiya, Go and Miki, Yohei and Mori, Masao and Nomoto, Ken'ichi},
  date = {2012-06-06},
  year = {2012},
  eprint = {1206.1199},
  eprinttype = {arXiv},
  eprintclass = {astro-ph},
  doi = {10.48550/arXiv.1206.1199},
  pubstate = {prepublished}
}

@article{losassoSpatiallyAdaptiveTechniques2006,
  title = {Spatially Adaptive Techniques for Level Set Methods and Incompressible Flow},
  author = {Losasso, Frank and Fedkiw, Ronald and Osher, Stanley},
  date = {2006-12-01},
  year = {2006},
  journaltitle = {Computers \& Fluids},
  journal = {Comput. \& Fluids},
  volume = {35},
  number = {10},
  pages = {995--1010},
  issn = {0045-7930},
  doi = {10.1016/j.compfluid.2005.01.006}
}

@article{mirzadehParallelLevelsetMethods2016,
  title = {Parallel Level-Set Methods on Adaptive Tree-Based Grids},
  author = {Mirzadeh, Mohammad and Guittet, Arthur and Burstedde, Carsten and Gibou, Frederic},
  date = {2016-10-01},
  year = {2016},
  journaltitle = {Journal of Computational Physics},
  journal = {J. Comput. Phys.},
  volume = {322},
  pages = {345--364},
  issn = {0021-9991},
  doi = {10.1016/j.jcp.2016.06.017}
}

@article{meagherGeometricModelingUsing1982,
  title = {Geometric Modeling Using Octree Encoding},
  author = {Meagher, Donald},
  date = {1982-06-01},
  year = {1982},
  journal = {Computer Graphics and Image Processing},
  volume = {19},
  number = {2},
  pages = {129--147},
  issn = {0146-664X},
  doi = {10.1016/0146-664X(82)90104-6}
}

@article{ghizzoLowHighfrequencyNature2020,
  title = {Low- and High-Frequency Nature of Oblique Filamentation Modes. {{II}}. {{Vlasov}}–{{Maxwell}} Simulations of Collisionless Heating Process},
  author = {Ghizzo, A. and Del Sarto, D.},
  date = {2020-07-21},
  year = {2020},
  journal = {Physics of Plasmas},
  volume = {27},
  number = {7},
  pages = {072104},
  issn = {1070-664X},
  doi = {10.1063/5.0003698}
}

@article{brizardFoundationsNonlinearGyrokinetic2007a,
  title = {Foundations of Nonlinear Gyrokinetic Theory},
  author = {Brizard, A. J. and Hahm, T. S.},
  date = {2007-04-02},
  year = {2007},
  journaltitle = {Reviews of Modern Physics},
  journal = {Rev. Mod. Phys.},
  volume = {79},
  number = {2},
  pages = {421--468},
  publisher = {American Physical Society},
  doi = {10.1103/RevModPhys.79.421}
}

@article{connorFastConstructionKnearest2010,
  title = {Fast Construction of K-Nearest Neighbor Graphs for Point Clouds},
  author = {Connor, Michael and Kumar, Piyush},
  date = {2010-07},
  year = {2010},
  journal = {IEEE Transactions on Visualization and Computer Graphics},
  volume = {16},
  number = {4},
  pages = {599--608},
  issn = {1941-0506},
  doi = {10.1109/TVCG.2010.9},
  eventtitle = {{{IEEE Transactions}} on {{Visualization}} and {{Computer Graphics}}}
}

@misc{bleifreiFreifrauvonbleifreiDyAda2025,
  title = {freifrauvonbleifrei/{{DyAda}}},
  author = {Pollinger, Theresa},
  date = {2025-06-12T14:10:08Z},
  origdate = {2024-06-12T06:07:17Z},
  year = {2025},
  note = {version 0.0.5},
  url = {https://github.com/freifrauvonbleifrei/DyAda},
  urldate = {2025-06-24}
}

@misc{bleifreiFreifrauvonbleifreiThingies_with_omnitrees2025,
  title = {freifrauvonbleifrei/thingies\_with\_omnitrees},
  author = {Pollinger, Theresa},
  date = {2025-06-12T14:10:08Z},
  year = {2025},
  url = {https://github.com/freifrauvonbleifrei/thingies_with_omnitrees},
  urldate = {2025-06-24},
}

@inproceedings{wohlerEstablishingDLRF25Research2024,
  title = {Establishing the {{DLR-F25}} as a Research Baseline Aircraft for the Short-Medium Range Market in 2035},
  author = {Wöhler, Sebastian and Häßy, Jannik and Kriewall, Vivian},
  date = {2024-09},
  year = {2024},
  publisher = {International Council of the Aeronautical Sciences},
  url = {https://www.icas.org/icas_archive/icas2024/data/papers/icas2024_0398_paper.pdf},
  eventtitle = {34th {{Congress}} of the {{International Council}} of the {{Aeronautical Sciences}}, {{ICAS}} 2024},
  langid = {english}
}

@misc{DLRF25,
  title = {{{DLR-F25}}},
  author = {{Deutsches Zentrum für Luft- und Raumfahrt e.V. (DLR)}},
  howpublished = {DLR Digital Hangar},
  url = {https://www.digital-hangar.de/portfolio/dlr-f25/},
  urldate = {2025-06-26}
}

@inproceedings{losassoSimulatingWaterSmoke2004,
  title = {Simulating Water and Smoke with an Octree Data Structure},
  booktitle = {{{ACM SIGGRAPH}} 2004 {{Papers}}},
  author = {Losasso, Frank and Gibou, Frédéric and Fedkiw, Ron},
  date = {2004-08-01},
  year = {2004},
  series = {{{SIGGRAPH}} '04},
  pages = {457--462},
  publisher = {Association for Computing Machinery},
  location = {New York, NY, USA},
  doi = {10.1145/1186562.1015745},
  isbn = {978-1-4503-7823-9}
}

@article{mirjaliliInterfacecapturingMethodsTwophase2017,
  title = {Interface-Capturing Methods for Two-Phase Flows: {{An}} Overview and Recent Developments},
  shorttitle = {Interface-Capturing Methods for Two-Phase Flows},
  author = {Mirjalili, Shahab and Jain, Suhas S. and Dodd, Micheal},
  date = {2017},
  year = {2017},
  journaltitle = {Center for Turbulence Research Annual Research Briefs},
  volume = {2017},
  number = {117--135},
  pages = {13},
  publisher = {Center for Turbulence Research},
  url = {https://doddm.com/publications/2017-ctr-sm-sj-md.pdf},
  urldate = {2025-07-23}
}

@inproceedings{vardakisUsingMulticompartmentalPoroelasticity2021,
  title = {Using {{Multicompartmental Poroelasticity}} to {{Explore Brain Biomechanics}} and {{Cerebral Diseases}}},
  booktitle = {Advances in {{Critical Flow Dynamics Involving Moving}}/{{Deformable Structures}} with {{Design Applications}}},
  author = {Vardakis, John C. and Guo, Liwei and Chou, Dean and Ventikos, Yiannis},
  editor = {Braza, Marianna and Hourigan, Kerry and Triantafyllou, Michael},
  date = {2021},
  pages = {151--163},
  publisher = {Springer International Publishing},
  location = {Cham},
  doi = {10.1007/978-3-030-55594-8_15},
  isbn = {978-3-030-55594-8},
  langid = {english}
}

@online{lyuNovelComputationalPreProcedural2024,
  title = {A {{Novel Computational Pre-Procedural Planning Model}} for {{Coronary Interventions Based}} on {{Coronary CT Angiography}}},
  author = {Lyu, Mengzhe and Liang, Ce and Zhang, Xuehuan and Wang, Xiao and Li, Qiaoqiao and Torii, Ryo and Ventikos, Yiannis and Chen, Duanduan},
  date = {2024-07-31},
  eprinttype = {bioRxiv},
  eprintclass = {New Results},
  pages = {2024.07.29.605713},
  doi = {10.1101/2024.07.29.605713},
  url = {https://www.biorxiv.org/content/10.1101/2024.07.29.605713v1},
  urldate = {2025-08-22},
  langid = {english},
  pubstate = {prepublished}
}

@article{peachVirtualComparisonECLIPs2019,
  title = {A {{Virtual Comparison}} of the {{eCLIPs Device}} and {{Conventional Flow-Diverters}} as {{Treatment}} for {{Cerebral Bifurcation Aneurysms}}},
  author = {Peach, T. W. and Ricci, D. and Ventikos, Y.},
  date = {2019-09-01},
  journaltitle = {Cardiovascular Engineering and Technology},
  shortjournal = {Cardiovasc Eng Tech},
  volume = {10},
  number = {3},
  pages = {508--519},
  issn = {1869-4098},
  doi = {10.1007/s13239-019-00424-3},
  url = {https://doi.org/10.1007/s13239-019-00424-3},
  urldate = {2025-08-22},
  langid = {english}
}

@dataset{pollingerReproducibilityDataBeauty2025,
  title = {Reproducibility {{Data}} for: {{The Beauty}} of {{Adaptive Mesh Refinement}}: {{Omnitrees}} for {{Efficient Dyadic Discretizations}}},
  shorttitle = {Reproducibility {{Data}} For},
  author = {Pollinger, Theresa},
  date = {2025-07-15},
  publisher = {Zenodo},
  doi = {10.5281/zenodo.15909872},
  url = {https://zenodo.org/records/15909872},
  urldate = {2026-02-26}
}

@inproceedings{musethOpenVDBOpensourceData2013,
  title = {{{OpenVDB}}: An Open-Source Data Structure and Toolkit for High-Resolution Volumes},
  shorttitle = {{{OpenVDB}}},
  booktitle = {{{ACM SIGGRAPH}} 2013 {{Courses}}},
  author = {Museth, Ken and Lait, Jeff and Johanson, John and Budsberg, Jeff and Henderson, Ron and Alden, Mihai and Cucka, Peter and Hill, David and Pearce, Andrew},
  date = {2013-07-21},
  series = {{{SIGGRAPH}} '13},
  pages = {1},
  publisher = {Association for Computing Machinery},
  location = {New York, NY, USA},
  doi = {10.1145/2504435.2504454},
  url = {https://dl.acm.org/doi/10.1145/2504435.2504454},
  urldate = {2026-02-25},
  isbn = {978-1-4503-2339-0}
}
}{
}

\end{document}